\documentclass[a4paper,UKenglish,cleveref, autoref, thm-restate]{lipics-v2021}

\pdfoutput=1 
\hideLIPIcs  

\usepackage[many]{tcolorbox}
\usepackage{mathtools}
\usepackage{amsmath}
\usepackage{bussproofs}
\usepackage{mathabx}
\usepackage{amssymb, amsthm}
\usepackage{stmaryrd}
\usepackage{float}
\usepackage{sfmath}
\tcbuselibrary{skins}

\newcommand{\dkbox}[2]{\begin{center}\begin{tcolorbox}[blanker,left=3mm,right=3mm,
  borderline vertical={2pt}{0pt}{black}, text width=#1mm, fontupper=\small]
\abovedisplayskip=0pt
  \begin{flalign*}
    #2
  \end{flalign*}
\end{tcolorbox}\end{center}}

\definecolor{darkblue}{RGB}{0,60,220}

\theoremstyle{plain}
\newtheorem{assumption}[theorem]{Assumption}
\newtheorem{convention}[theorem]{Convention}

\newcommand{\ctb}[1]{\textcolor{darkblue}{#1}}
\newcommand{\Type}{\text{\textup{{\texttt{TYPE}}}}}
\newcommand{\WF}{\textup{well-formed}}

\newcommand{\El}[1]{\ctb{El_{#1}}}
\newcommand{\app}[2]{\ctb{app_{#1,#2}}}
\newcommand{\betarule}[2]{\ctb{beta_{#1,#2}}}
\newcommand{\abs}[2]{\ctb{abs_{#1,#2}}}
\newcommand{\Prod}[2]{\ctb{Prod_{#1,#2}}}
\newcommand{\U}[1]{\ctb{U_{#1}}}

\newcommand{\Kind}{\text{\textup{{\texttt{KIND}}}}}

\newcommand{\Agda}{\textsc{Agda}}
\newcommand{\Coq}{\textsc{Coq}}
\newcommand{\ELF}{\textsc{LF}}

\newcommand{\Dedukti}{\textsc{Dedukti}}
\newcommand{\Lambdapi}{\textsc{Lambdapi}}

\newcommand{\red}{\xhookrightarrow{\quad}}
\newcommand{\invred}{\xhookleftarrow{\quad}}
\newcommand{\uu}[1]{\ctb{u_{#1}}}

\newcommand{\invt}[1]{|#1|}
\newcommand{\trans}[1]{\llbracket #1 \rrbracket}
\newcommand{\tytrans}[1]{REMOVE-ME}
\newcommand{\bind}[1]{{\scriptstyle [#1]}}
\newcommand{\cont}[1]{|#1|}

\newcommand{\tycont}[1]{\lVert #1 \rVert}
\newcommand{\vdashdk}{\vdash_\textup{\texttt{DK}}}
\usepackage{thmtools}

\newcommand{\thref}[1]{{\hypersetup{hidelinks}\textit{\nameref{#1}}} (\autoref{#1})}
\newcommand{\tolong}[1]{#1}
\newcommand{\toshort}[1]{}
\newcommand{\Lambdadk}{\Lambda_{\textup{\texttt{DK}}}}
\newcommand{\Sigmaepts}{\Sigma_{\textup{\texttt{EPTS}}}}
\newcommand{\Repts}{\mathscr{R}_{\textup{\texttt{EPTS}}}}

\title{Adequate and computational encodings in the logical framework Dedukti}

\author{Thiago Felicissimo}{Université Paris-Saclay, INRIA project
Deducteam, Laboratoire de Méthodes Formelles, ENS Paris-Saclay, 91190 France}{thiago.felicissimo@inria.fr}{}{}

\authorrunning{T. Felicissimo} 

\Copyright{Thiago Felicissimo} 



\category{} 

\tolong{\relatedversion{This is a long version of a paper accepted at FSCD 2022.}}
\toshort{\relatedversion{}} 

\acknowledgements{I would like to thank my PhD advisors Frédéric Blanqui and Gilles Dowek for the helpful discussions and comments on this paper. I would also like to thank the anonymous reviewers for their very helpful comments and suggestions. }

\nolinenumbers 

\EventEditors{Amy P. Felty}
\EventNoEds{1}
\EventLongTitle{7th International Conference on Formal Structures for Computation and Deduction (FSCD 2022)}
\EventShortTitle{FSCD 2022}
\EventAcronym{FSCD}
\EventYear{2022}
\EventDate{August 2--5, 2022}
\EventLocation{Haifa, Israel}
\EventLogo{}
\SeriesVolume{228}
\ArticleNo{21}

\begin{document}

\maketitle

\begin{abstract}
  \Dedukti~is a very expressive logical framework which unlike most frameworks, such as the Edinburgh Logical Framework (\ELF), allows for the representation of computation alongside  deduction. However, unlike \ELF{} encodings, \Dedukti{} encodings proposed until now do not feature an adequacy theorem --- \textit{i.e.}, a bijection between terms in the encoded system and in its encoding. Moreover, many of them also do not  have a conservativity result, which compromises the ability of \Dedukti~to check proofs written in such encodings. We propose a different approach for \Dedukti~encodings which do not only allow for simpler conservativity proofs, but which also restore the adequacy of encodings. More precisely, we propose in this work adequate (and thus conservative) encodings for Functional Pure Type Systems. However, in contrast with \ELF~encodings, ours is computational --- that is, represents computation directly as computation. Therefore, our work is the first to present and prove correct an approach allowing for encodings that are both adequate and computational in~\Dedukti.

  \end{abstract}

\section{Introduction}
\label{sec:intro}

The research on  proof checking naturally leads to the proposal of many logical systems and theories. \textit{Logical frameworks} are a way of addressing this heterogeneity by proposing a common foundation in which systems and theories can be defined. The \textit{Edinburgh Logical Framework} (\ELF)\cite{ELF} is one of the milestones in the history of logical frameworks, and proposes the use of a dependently-typed lambda-calculus to express deduction. However, as modern proof assistants move from traditional logics to type theories, where computation plays an important role alongside deduction, it becomes essential for such frameworks to also be able to express computation, something that the \ELF~does not achieve.

The logical framework \Dedukti\cite{dedukti} addresses this point by extending the \ELF~with rewriting rules, thus allowing for the representation of both deduction and computation. This framework was already proven to be as a very expressive system, and has been used to encode the logics of many proof assistants, such as \Coq\cite{gaspard}, \Agda\cite{genestier}, \textsc{PVS}\cite{proofIrel} and others.

However, an unsatisfying aspect persists as, unlike \ELF{} encodings, the \Dedukti{} encodings proposed until now are  not \textit{adequate}, in the sense that they do not feature a syntactical bijection between the terms of the encoded system and those of the encoding.  Such property is key to ensure that the framework faithfully represents the syntax on the encoded system. Moreover, proving that \Dedukti{} encodings are \textit{conservative} (\textit{i.e.}, that if the translation of a type is inhabited, then this type is inhabited) is still a challenge, in particular for recent works such as \cite{gaspard}\cite{thire}\cite{genestier}\cite{proofIrel}.  This is a problem if one intends to use \Dedukti~to check the correctness of proofs coming from proof assistants: if  conservativity does not hold then the fact that the translation of a proof is checked correct in \Dedukti~does not imply that this proof is correct.

In the specific case of Pure Type Systems (PTS), a class of type systems which generalizes many others, \cite{dowek2007} was the first to propose an encoding of functional PTSs into \Dedukti{}. One of their main contributions is that, contrarily to \ELF{} encodings, theirs is \textit{computational} --- that is, represents computation in the encoded system directly as computation. The authors then show that the encoding is  conservative under the hypothesis of normalization of their rewrite rules.

To address the issue of this unproven assumption, \cite{dowekmodels} proposed a  notion of model of \Dedukti~and showed using the technique of \textit{reducibility candidates} that the existence of such a model entails the normalization of the encoding. Using this result, the author then showed the conservativity of the encoding of Simple Type Theory and of the Calculus of Constructions. This technique however is not very satisfying as the construction of such models is a very technical task, and needs to be done case by case. One can also wonder why conservativity should rely on normalization.

The cause of this difficulty in \cite{dowek2007} and in all other traditional \Dedukti{} encodings comes from a choice   made to represent the abstraction and application of the encoded system directly by the abstraction and application of the framework. This causes a confusion as redexes of the encoded system, that represent real computations, get confused with the $ \beta $ redexes of the framework, which in other frameworks such as the \ELF{}  are used exclusively to represent binder substitution. As a non-normal term can contain both types of redexes, it is impossible to inverse translate it as some of these redexes are ill-typed in the original system, and the only way of eliminating these ill-typed redexes is by reducing all of them. One then needs this process to be terminating, which is non-trivial to show as it involves proving that the reduction of the redexes of the encoded system terminates.

The work of \cite{assaf2015conservativity} first noted this problem and proposed a different approach to show the conservativity of the encoding of PTSs. Instead of relying on the normalization of the encoding, they proposed to directly inverse translate  terms without normalizing them. As this creates ill-typed terms, they then used reducibility candidates to show that these ill-typed terms reduce to well-typed ones, thus proving conservativity for the encoding in \cite{dowek2007}.

Even though this technique is a big improvement over \cite{dowekmodels}, it is still unsatisfying that both of them rely on involved arguments using reducibility, whereas the proofs of \ELF~encodings were very natural. They also both rely on intricate properties of the encoded systems, which is unnatural given that logical frameworks should ideally only require the encoded systems to satisfy some basic properties, and be agnostic with respect to more deep ones --- for instance, one should not be obliged to show that a given system is consistent in order to encode it in a logical framework. This reason, coupled with the technicality of these proofs, may explain why recent works such as \cite{thire}, \cite{proofIrel} and \cite{genestier} have left conservativity as conjecture. Moreover, none of these works have addressed the lack of an adequacy theorem, which until now has remained an overlooked problem in the \Dedukti{} community.

\subsection*{Our contribution}
\label{subsec:contribution}

We propose to depart from the approach of traditional \Dedukti~encodings by restoring the separation that existed in \ELF~encodings. Our paradigm  represents the abstractions and applications of the encoded system not by those of the framework, but by dedicated constructions. Using this approach, we propose an encoding of functional PTSs that is not only sound and conservative but also adequate. However, in contrast with \ELF~encodings, ours is computational like other \Dedukti~encodings.

To show conservativity, we leverage the fact that the computational rules of the encoded system are not represented by $ \beta $ reduction anymore, but by dedicated rewrite rules. This allows us to normalize only the framework's $ \beta $ redexes without touching those associated with the encoded system, and thus  performing no computation from its point of view.

To be able to $ \beta$ normalize terms, we generalize the proof in \cite{ELF} to give a general criterion for the normalization of $ \beta $ reduction in \Dedukti. This criterion imposes rewriting rules to be \textit{arity preserving} (a definition we introduce). This is not satisfied by traditional \Dedukti~encodings, but poses no problem to ours. The proof uses the simple technique of defining  an  erasure map into the simply-typed lambda calculus, which is known to be normalizing.

\subsection*{Outline}
\label{subsec:outline}

We start in Section \ref{sec:dedukti} by recalling the preliminaries about \Dedukti{}. We proceed in Section \ref{sec:sn_beta} by proposing a criterion for the normalization of $ \beta $ in \Dedukti, which is used in our proofs of conservativity and adequacy. In Section \ref{sec:pts} we introduce an explicitly-typed version of Pure Type Systems, which is used for the encoding.  We then present the encoding in Section \ref{sec:encoding}, and proceed by showing it is sound in Section \ref{sec:soundness} and that it is conservative and adequate in Section \ref{sec:conservativity}.  In Section \ref{sec:representation_sorts} we discuss how our approach can be used together with already known techniques to represent systems with infinitely many sorts. Finally, in Section \ref{sec:practice} we discuss more practical aspects by showing how the encoding can be instantiated and used in practice.

\section{Dedukti}
\label{sec:dedukti}

\begin{figure}[h]
{\small
\begin{center}
  \AxiomC{}
\RightLabel{\texttt{Empty}}
\UnaryInfC{$\Sigma; -~\texttt{well-formed}$}
\DisplayProof
\hskip 1.5em
\AxiomC{$\Sigma;\Gamma \vdash A : \Type$}
\RightLabel{\texttt{Decl}}
\LeftLabel{$x \notin \Gamma$}
\UnaryInfC{$\Sigma;\Gamma, x : A~\texttt{well-formed} $}
\DisplayProof
\end{center}
\begin{center}
\AxiomC{$\Sigma;\Delta \vdash A : s $}
\AxiomC{$\Sigma;\Gamma \vdash \vec{M} : \Delta $}  
\RightLabel{\texttt{Cons}}
\LeftLabel{$c [\Delta] : A \in \Sigma$}
\BinaryInfC{$\Sigma;\Gamma \vdash c[\vec{M}] : A\{\vec{M}\} $}
\DisplayProof
\hskip 1.5em
  \AxiomC{$ \Sigma;\Gamma~\texttt{well-formed} $}
\RightLabel{\texttt{Sort}}
\UnaryInfC{$\Sigma;\Gamma \vdash \Type : \Kind$}
\DisplayProof
\end{center}
\begin{center}
\AxiomC{$\Sigma;\Gamma~\texttt{well-formed}$}
\RightLabel{\texttt{Var}}
\LeftLabel{$ x : A \in \Gamma $}
\UnaryInfC{$\Sigma;\Gamma \vdash x : A  $}
\DisplayProof
\hskip 3em
  \AxiomC{$\Sigma;\Gamma \vdash M : A $}
  \AxiomC{$\Sigma;\Gamma \vdash B : s $}  
  \RightLabel{\texttt{Conv}}
  \LeftLabel{$A \equiv_{\beta\mathscr{R}} B$}  
\BinaryInfC{$\Sigma;\Gamma \vdash M : B $}
\DisplayProof
\end{center}
\begin{center} 
  \AxiomC{$\Sigma; \Gamma \vdash A : \Type $}
  \AxiomC{$\Sigma; \Gamma, x : A \vdash B : s $}  
\RightLabel{\texttt{Prod}}
\BinaryInfC{$\Sigma;\Gamma \vdash \Pi    x : A . B : s  $}
\DisplayProof
\hskip 1.5em
\AxiomC{$\Sigma;\Gamma \vdash M : \Pi x : A . B  $}
\AxiomC{$\Sigma; \Gamma \vdash N : A $}
\RightLabel{\texttt{App}}
\BinaryInfC{$\Sigma;\Gamma \vdash M N : B\{N/x\} $}
\DisplayProof
\end{center}
\begin{center}
  \AxiomC{$\Sigma; \Gamma \vdash A : \Type $}
  \AxiomC{$\Sigma; \Gamma, x : A \vdash B : s $}  
  \AxiomC{$\Sigma; \Gamma, x : A \vdash M : B $}
  \RightLabel{\texttt{Abs}}
\TrinaryInfC{$\Sigma;\Gamma \vdash \lambda x : A . M :\Pi x : A . B  $}
\DisplayProof
\end{center}}
\caption{Typing rules for \Dedukti}
\label{typing-dk}
\end{figure}

The logical framework \Dedukti{} \cite{dedukti}~has the syntax of the $ \lambda $-calculus with dependent types \cite{ELF} ($ \lambda\Pi $-calculus). Like works such as \cite{proofIrel}, we consider here a version with arities, with the following syntax.
\begin{align*}
  A, B, M, N &::= x \mid c[\vec{M}]~|~\Type~|~\Kind~|~M N~|~\lambda x : A . M~|~\Pi x : A. B
\end{align*}
Here, $ c $ ranges in an infinite set of constants $ \mathcal{C} $, and $ x $ ranges in an infinite set of variables $ \mathcal{V} $. Each constant $ c $ is assumed to have a fixed arity $ n_c $ and for each occurence of $ c[\vec{M}] $ we should have $ length(\vec{M})=n_c $. We denote $ \Lambdadk $ the set of terms generated by this grammar. We call a term of the form $ \Pi x : A.B $ a \textit{dependent product}, and we write $ A \to B $ when $ x $ does not appear free in $ B $. We allow ourselves sometimes to write $ c~\vec{M} $ instead of $ c[\vec{M}] $ to ease the notation.

A \textit{context} $ \Gamma $ is a finite sequence of pairs $ x : A $ with $ A \in \Lambdadk$. A \textit{signature} $ \Sigma $ is a finite set of triples $ c[\Delta] : A $ where $ A \in \Lambdadk $ and $ \Delta$ is a context containing at least all free variables of $ A$. The main difference between \Dedukti~and the $ \lambda \Pi $-calculus is that we also consider a set $ \mathscr{R} $ of \textit{rewrite rules}, that is, of pairs of the form $ c[\vec{l}] \red r $ with $ l_1,...,l_k,r \in \Lambdadk $. A \textit{theory} is a pair $ (\Sigma, \mathscr{R}) $ such that all constants appearing in $ \mathscr{R} $ are declared in $ \Sigma $.

We  write $ \red_\mathscr{R} $ for the context and substitution closure of the rules in $ \mathscr{R} $ and $ \red_{\beta\mathscr{R}} $ for $ \red_\beta \cup \red_\mathscr{R} $. We also consider the equivalence relation $ \equiv_{\beta\mathscr{R}} $ generated by $ \red_{\beta\mathscr{R}} $. Finally, we may refer to $ \red_{\beta\mathscr{R}} $ and $ \equiv_{\beta\mathscr{R}} $ by just $ \red$ and $ \equiv $.

Typing in \Dedukti{} is given by the rules in Figure \ref{typing-dk}. In rule \texttt{Cons} we use the usual notation $ \Sigma; \Gamma \vdash \vec{M} : \Delta $ meaning that $ \Delta=x_1:A_1,...,x_n:A_n$ and $ \Sigma; \Gamma \vdash M_i : A_i\{M_1/x_1\}...\{M_{i-1}/x_{i-1}\} $ is derivable for $ i = 1,...,n $. We then also allow ourselves to write $ A\{\vec{M}\} $ instead of $ A\{M_1/x_1\}...\{M_n/x_n\} $. \toshort{We refer to Appendix \ref{sec:meta_dk} for a review of some basic  metatheorems.}

\tolong{
  We recall the following basic metatheorems.

  \begin{proposition}[Basic properties]
  Suppose $ \red_{\beta\mathscr{R}} $ is confluent.
  \begin{enumerate}
  \item Weakening: If $ \Sigma; \Gamma \vdash M : A $, $ \Gamma \sqsubseteq \Gamma' $ and $ \Sigma;\Gamma'~\textup{\texttt{well-formed}} $ then $ \Sigma;\Gamma' \vdash M : A $
  \item Well-typedeness of contexts: If $ \Sigma;\Gamma~\textup{\texttt{well-formed}} $ then for all $x : B \in \Gamma $, $ \Sigma;\Gamma \vdash B : \Type $
  \item Inversion of typing: Suppose $ \Sigma;\Gamma \vdash M : A $
    \begin{itemize}
    \item If $ M = x $ then $ x : A' \in \Gamma $ and $ A \equiv A' $
    \item If $ M = c[\vec{N}] $ then $ c[\Delta] : A'  \in \Sigma $, $ \Sigma;\Delta \vdash A' : s $, $ \Sigma;\Gamma \vdash \vec{N} : \Delta $ and $ A'\{\vec{N}/\Delta\} \equiv A $
    \item If $ M = \Type $ then $ A \equiv \Kind $
    \item  $ M= \Kind $ is impossible
    \item If $ M = \Pi x : A_1. A_2 $ then $ \Sigma;\Gamma \vdash A_1 : \Type $, $ \Sigma; \Gamma,x:A_1 \vdash A_2 : s $ and $ s \equiv A $
    \item If $ M = M_1 M_2 $ then $ \Sigma; \Gamma \vdash M_1 : \Pi x: A_1.A_2 $, $ \Sigma;\Gamma \vdash M_2 : A_1 $ and $ A_2\{M_2/x\} \equiv A $
    \item If $ M = \lambda x : B. N $ then $ \Sigma;\Gamma \vdash B : \Type $, $ \Sigma; \Gamma, x:B \vdash C:s $, $ \Sigma;\Gamma,x:B \vdash N:C $ and $ A \equiv \Pi x:B.C $
    \end{itemize}
  \item Uniqueness of types: If $ \Sigma;\Gamma \vdash M : A $ and $ \Sigma;\Gamma \vdash M : A' $ then $ A \equiv A' $
  \item Well-sortness: If $ \Sigma;\Gamma \vdash M : A $ then $ \Sigma;\Gamma \vdash A : s $ or $ A = \Kind $
  \end{enumerate}
\end{proposition}

\begin{theorem}[Conv in context for DK]
  \label{dk_conv_in_context}
  Let $ A \equiv A' $ with $ \Sigma;\Gamma \vdash A' : s $. We have
  \begin{itemize}
  \item $ \Sigma;\Gamma, x : A, \Gamma'~\textup{\texttt{well-formed}} \Rightarrow \Sigma;\Gamma, x : A', \Gamma'~\textup{\texttt{well-formed}}$
  \item $ \Sigma;\Gamma, x : A, \Gamma' \vdash M : B \Rightarrow \Sigma;\Gamma, x : A', \Gamma' \vdash M : B $
  \end{itemize}
\end{theorem}

\begin{proposition}[Reduce type in judgement]
  \label{reduce_type}
  Suppose $ \red_{\beta\mathscr{R}} $ is confluent and satisfies subject reduction. Then if $ \Sigma; \Gamma \vdash M : A $ and $ A \red^* A' $ we have $ \Sigma; \Gamma \vdash M : A' $.
\end{proposition}
}

\tolong{
\section{Strong Normalization of $ \beta $ in Dedukti}
\label{sec:sn_beta}

In order to show the conservativity of encodings, one often needs to $ \beta $ normalize  terms, thus requiring $ \beta $ to be normalizing for well-typed terms. In this section we generalize the proof of normalization of the $ \lambda \Pi $-calculus given in \cite{ELF}  to \Dedukti{}. More precisely, we show that, given that $ \beta\mathscr{R} $ is confluent and \textit{arity preserving} (a definition we will introduce in this section), then $ \beta $ is SN (strongly normalizing) in \Dedukti{} for well-typed terms.

Note that, unlike works such as \cite{guillaume-termination}, which provide syntactic criteria on the normalization of $ \beta\mathscr{R} $ in \Dedukti, we only aim to show the normalization of $ \beta $. In particular $ \beta\mathscr{R} $ may not be SN in our setting. Our work has more similar goals to \cite{relevance-of-proof-irrelevance}, which provides criteria for the SN of $ \beta $ in the Calculus of Constructions when adding object-level rewrite rules. However, our work also allows for type-level rewrite rules, which will be needed in our encoding. 

Our  proof works by defining an erasure map into the simply-typed $ \lambda $-calculus, which is known to be SN, and then show that this map preserves typing and non-termination of $ \beta $, thus implying that $ \beta $ is SN in \Dedukti. To do this, the erasure map must remove the dependency inside types, but as in \Dedukti~terms and types are all mixed together, we will first need to be able to separate them syntactically.

The syntactic stratification theorem (Theorem \ref{stratification}) is a standard property of \Dedukti~and is exactly what we need here. However, unlike the known variants in the literature, such as in \cite{frederic-phd}, we prove a more general version not requiring subject reduction of $ \beta\mathscr{R} $. The proof draws inspiration from a similar one in \cite{relevance-of-proof-irrelevance}.

We  proceed as follows. First we start by proving our generalization of the stratification theorem (Thereon \ref{stratification}). This is followed by the definition of the erasure function from \Dedukti{} into the simply-typed $ \lambda $-calculus (Definition \ref{erasure-map}).  We then introduce our definition of \textit{arity preserving} rewrite systems (Definition \ref{arity-preserving}), and give some motivation of why the proof works. We then show that this function preserves both typing (Proposition \ref{pres_typing_cont}) and non-termination (Proposition \ref{pres_beta_cont}). Finally, by putting all this together, we will conclude by showing our main result (Theorem \ref{norm_beta_dk}).

\subsection{Syntactic stratification}
\label{subsec:strat}

\begin{definition}
We  introduce the following basic definitions.
\begin{enumerate}
\item Given a signature $ \Sigma $, a constant $ c $ is  type-level (and referred by $ \alpha,\gamma$) if  $ c[\Delta] :A \in\Sigma $ with $ A $ of the form $ \Pi \vec{x}:\vec{B}:\Type $, otherwise it is  object-level (and referred by $ a,b$).
\item A rewrite rule $ c[\vec{l}] \red r $ is type-level if its head symbol $ c $ is a type-level constant.
\end{enumerate}
\end{definition}

As previously mentioned, our proof of the stratification theorem will not need subject reduction of $ \mathscr{R} $. Instead, we will only need the following syntactic property.

\begin{definition}
We say that $ \mathscr{R} $ is well-formed (with respect to $ \Sigma $) if for all type-level rules, its right-hand side is in the following grammar, where $ \vec{M}, N, B $ are arbitrary.
  {\small  \begin{align*}
           R &::= \alpha[\vec{M}] \mid R N \mid \lambda x : B. R \mid \Pi x : R. R 
  \end{align*}}
\end{definition}

We are now ready to define the syntactic classes of terms in \Dedukti{}, through the following grammars. We will show in Theorem \ref{stratification} that every typed term belong to one of these classes.
{\small\begin{align*}
  K &::= \Type \mid \Pi x : T. K  & \text{(Kinds)}\\
  T &::= \alpha[\vec{O}] \mid T O \mid \lambda x : T. T \mid \Pi x : T. T  & \text{(Type Families)}\\
  O &::= x \mid a[\vec{O}] \mid O O \mid \lambda x : T. O  & \text{(Objects)}  
\end{align*}}
These grammars can easily be shown closed under object-level substitution.

\begin{lemma}[Closure under object-level substitution]
  For all objects $ O $, we have
  \begin{itemize}
  \item If $ M $ is an object, then $ M\{O/x\} $ is an object
  \item If $ M $ is an type family, then $ M\{O/x\} $ is a type family
  \item If $ M $ is a kind, then $ M\{O/x\} $ is a kind
  \end{itemize}
\end{lemma}

A property that one would find natural is for these grammars to be closed under reduction. However, as it is shown by the following example, this is not the case.

\begin{example}
Consider the rule $ \alpha[\lambda x : y. x] \red \alpha[y] $. With the substitution $ y \mapsto \gamma $ we have $ \alpha[\lambda x : \gamma. x] \red \alpha[\gamma] $. The left-hand side is a type family, whereas the right-hand side is not in any of the grammars.
\end{example}

However, we can still define a weaker notion of pre-kinds and pre-type families for which closure under rewriting holds. This property will be key when proving the stratification theorem.

\begin{lemma}
  Define the grammars
  {\small  \begin{align*}
           L &::= \Type \mid \Pi x : R. L  & \text{(Pre-Kinds)}           \\             
           R &::= \alpha[\vec{M}] \mid RN \mid \lambda x : A. R \mid \Pi x : R. R & \text{(Pre-Type Families)}
  \end{align*}}where $ A, N, \vec{M} $ are arbitrary.  If $ \mathscr{R} $ is well-formed, then they are disjoint and closed under $ \beta\mathscr{R} $.
\end{lemma}
\begin{proof}
The two grammars are clearly disjoint. Before showing closure under rewriting, we first show closure under substitution: for every pre-kind $ L $, pre-type family $ R $ and term $ N $, $ L \{ N/x \} $ is a pre-kind and $ R \{ N/x \} $ is a pre-type family (by induction on $ R $ and $ L $). Then, by induction on the rewrite context and using closure under substitution we show the result.
\end{proof}

We are now ready to show the stratification theorem.

\begin{theorem}[Syntactical stratification]
  \label{stratification}
  Suppose that $ \beta\mathscr{R} $ is confluent and $ \mathscr{R} $ is well formed. If  $ \Sigma;\Gamma \vdash M : A $ then exactly one of the following hold:
\begin{enumerate}
\item $ M$ is a kind and $ A = \Kind $
\item $ M $ is a type family and $ A $ is a kind
\item $ M $ is an object and $ A $ is a type family
\end{enumerate}
\end{theorem}
\begin{proof}
  First note that the grammars are clearly disjoint, so only one of the cases can hold. We proceed by showing the rest by induction over $ \Sigma;\Gamma \vdash M : A $.

\textbf{Sort}: Trivial.

  \textbf{Var}: We have
    \begin{center} 
  \AxiomC{$\Sigma; \Gamma~\texttt{well-formed} $}
  \LeftLabel{$x:A \in \Gamma $}
\RightLabel{\texttt{Var}}
\UnaryInfC{$\Sigma;\Gamma \vdash x : A  $}
\DisplayProof
\end{center}
For some $ \Gamma' \sqsubseteq \Gamma $, we have $ \Sigma;\Gamma' \vdash A : \Type $ with a smaller derivation tree. By IH, $ A $ is a type family, hence the result follows.

\textbf{Cons}: We have

\begin{center}
\AxiomC{$\Sigma;\Delta \vdash A : s $}
\AxiomC{$\Sigma;\Gamma \vdash \vec{M} : \Delta $}  
\RightLabel{\texttt{Cons}}
\LeftLabel{$c [\Delta] : A \in \Sigma$}
\BinaryInfC{$\Sigma;\Gamma \vdash c[\vec{M}] : A\{\vec{M}\} $}
\DisplayProof
\end{center}

We first show the following claim.

\begin{claim}
$ \vec{M} $ is made of objects.
\end{claim}
\begin{claimproof}
  Write $ \Delta = x_1: A_1,...,x_n : A_n $. First note that $ \Sigma; \Delta \vdash A : s $ implies  $ \Sigma; x_1: A_1,...,x_{i-1}:A_{i-1} \vdash A_i : \Type$ with a smaller derivation tree, hence by the IH each $ A_i $ is a type family. We now show by induction on $ i $ that for $ i = 1,...,n $,  $ M_i $ is an object\footnote{We will use the terms ``inner IH'' for the IH corresponding to this claim and ``outer IH'' for the IH corresponding to the whole thoerem.}.

  For the case $ i=1 $ this follows from $ \Sigma;\Gamma \vdash M_1 : A_1  $, by the outer IH and the fact that $ A_1 $ is a type family. For the induction step, we have $ \Sigma;\Gamma \vdash M_i: A_i \{M_1/x_1 \}...\{M_{i-1}/x_{i-1}\} $. We know that $ A_i $ is a type family, and by the inner IH we have that $ M_1,...,M_{i-1} $ are objects. By closure under object-level substitution, $ A_i \{M_1/x_1\}...\{M_{i-1}/x_{i-1}\} $ is also a type family. Hence the outer IH implies that $ M_i $ is an object.
\end{claimproof}

We now proceed with the main proof obligation. If $ s = \Kind $ by IH $ A $ is a kind, of the form $ \Pi \vec{x}:\vec{B}.\Type$. Hence $ c $ is a type-level constant.  Because $ c $ is type-level and $ \vec{M} $ is made of objects, then $ c[\vec{M}] $ is a type family. Finally, as $ \vec{M} $ are objects, then by closure under object-level substitution $ A \{\vec{M}\} $ is a kind. Hence we are in case $ 2 $.

If $ s = \Type $ by IH $ A $ is a type family. Hence $ c $ is an object-level constant. Because $ c $ is object-level and $ \vec{M} $ is made of objects, then $ c[\vec{M}] $ is an object. Finally, as $ \vec{M} $ are objects, then by closure under object-level substitution $ A \{\vec{M}\} $ is a type family. Hence we are in case $ 3 $.

  \textbf{Conv}: We have

  \begin{center} 
  \AxiomC{$\Sigma; \Gamma \vdash M : A $}
  \AxiomC{$\Sigma; \Gamma \vdash B : s $}
  \LeftLabel{$A \equiv B $}    
\RightLabel{\texttt{Conv}}
\BinaryInfC{$\Sigma;\Gamma \vdash M : B  $}
\DisplayProof
\end{center}

By confluence, there is $ C $ with $ A \red^* C \invred^* B $. Note that type families and kinds are also respectively pre-type families and pre-kinds, which are disjoint and closed under rewriting. Therefore, an important remark is that both situations in which $ A $ is a type-family and $ B $ a kind, or $ A $ a kind and $ B $ a type family, are impossible.

We now proceed with the proof and consider the cases $ s = \Type $ and $ s = \Kind $.

If $ s = \Type $ then $ B $ is a type family. Applying the IH to $ M : A $, then by the previous remark we only need to consider the case in which $ A $ is a type family, and thus $ M $ is an object as required.

If $ s = \Kind $, then $ B $ is a kind. Applying the IH to $ M : A $, then by previous remark we only need to consider the case in which $ A $ is a kind, and thus $ M $ is a type family as required.

  \textbf{Prod}: We have
  \begin{center} 
  \AxiomC{$\Sigma; \Gamma \vdash A : \Type $}
  \AxiomC{$\Sigma; \Gamma, x : A \vdash B : s $}  
\RightLabel{\texttt{Prod}}
\BinaryInfC{$\Sigma;\Gamma \vdash \Pi x : A . B : s  $}
\DisplayProof
\end{center}
We have either $ s= \Type $ or $ s = \Kind $.

If $ s = \Type $, then by IH both  $ A, B $ are type families, hence $ \Pi x : A. B $ is a type family and we are in case (2).

If $ s = \Kind $, then $ A $ is a type family and $ B $ a kind, hence $ \Pi x : A. B $ is a kind and we are in case (1).

  \textbf{Abs}: We have
\begin{center}
  \AxiomC{$\Sigma; \Gamma \vdash A : \Type $}
  \AxiomC{$\Sigma; \Gamma, x : A \vdash B : s $}  
  \AxiomC{$\Sigma; \Gamma, x : A \vdash M : B $}
  \RightLabel{\texttt{Abs}}
\TrinaryInfC{$\Sigma;\Gamma \vdash \lambda x : A . M :\Pi x : A . B  $}
\DisplayProof
\end{center}
We have either $ s= \Type $ or $ s = \Kind $.

If $ s = \Type $, then by IH both  $ A, B $ are type families and $ M $ is an object. Hence $ \lambda x : A. M $ is an object, $ \Pi x : A. B$ is a type family and we are in case (3).

If $ s = \Kind $, then $ B $ is a kind and $ A, M $ are type families. Hence $ \Pi x : A. M $ is a type family, $ \Pi x : A. B $ is a kind and we are in case (2).

\textbf{App}: We have
\begin{center}
  \AxiomC{$\Sigma;\Gamma \vdash M : \Pi x : A . B  $}
  \AxiomC{$\Sigma; \Gamma \vdash N : A $}
  \RightLabel{\texttt{App}}
\BinaryInfC{$\Sigma;\Gamma \vdash M N : B \{N/x\} $}
\DisplayProof
\end{center}
By the IH applied to $ M : \Pi x : A. B $, $ \Pi x : A. B $ is either a type family or a kind, hence in all cases $ A $ is a type family, and thus by the IH applied to $ N : A $, $ N $ is an object.

If $ \Pi x : A. B $ is a type family, then $ M $ is an object, and thus $ M N $ is also. As $ B $ is a type family and the grammars are closed by object-level substitution, $ B \{N/x \} $ is also a type family. Hence we are in case (2).

If $ \Pi x : A. B $ is a kind, then $ M $ is a type family, and thus $ M N $ is also. As $ B $ is a kind and the grammars are closed by object-level substitution, $ B \{N/x \} $ is a kind. Hence we are in case (1).
\end{proof}

\subsection{Erasure map}

We are now ready to give the definition of the erasure map into the simply-typed $ \lambda $-calculus.

\begin{definition}[Erasure map]\label{erasure-map}
  Consider the simple types generated by the  grammar \[
\sigma ::= * \mid \sigma \to \sigma
    \,.\]Moreover, let  $ \Gamma_\pi $ be the context containing for each $ \sigma $ the declaration $ \pi_\sigma : * \to (\sigma \to *) \to *$. We define the partial functions $ \tycont{-}, \cont{-} $ by the following equations.
  
{\small\noindent\parbox{.4\textwidth}{
\begin{align*}
 \tycont{\Type} &= *\\
 \tycont{\alpha[\vec{M}]} &= *\\
 \tycont{\Pi x : A. B} &= \tycont{A} \to \tycont{B}\\
 \tycont{A N} &= \tycont{A}\\
 \tycont{\lambda x : A. B} &= \tycont{B}
\end{align*}}
\parbox{.4\textwidth}{
\begin{align*}
\cont{x} &= x\\
\cont{a[\vec{M}]} &= a~|\vec{M}|\\
\cont{\alpha[\vec{M}]} &= \alpha~|\vec{M}|\\  
\cont{M N} &= \cont{M} \cont{N}\\
\cont{\lambda x : A. M} &= (\lambda z. \lambda x. \cont{M})\cont{A} \text{ where }z \notin FV(M)\\
\cont{\Pi x : A. B} &= \pi_{\tycont{A}}~\cont{A}~(\lambda x. \cont{B})
\end{align*}}}%
\end{definition}

In particular, note that $ \cont{-} $ is defined for all objects and type families, and that $ \tycont{-} $ is defined for all type-families and kinds. We also extend the definition  of $ \tycont{-} $ (partially) on contexts and signatures by the following equations.
\begin{align*}
  \tycont{-} &= -\\
  \tycont{x : A, \Gamma} &= x : \tycont{A}, \tycont{\Gamma}\\
\tycont{c[x_1 : A_1,...,x_n:A_n] : A; \Sigma} &= (c : \tycont{A_1} \to ... \to \tycont{A_n} \to \tycont{A}), \tycont{\Sigma}
\end{align*}

In order to show the normalization of $ \beta $, we need the erasure to preserve typing. The main obstacle when showing this is dealing with the \texttt{Conv} rule. To make the proof go through, we would need to show that  if  $ A \equiv B $ then $ \tycont{A}=\tycont{B} $. In the $ \lambda \Pi $-calculus this can be easily shown, however because in \Dedukti{} the relation $ \equiv $ also  takes into account the rewrite rules in $ \mathscr{R} $, we can easily build counterexamples in which this does not hold.

\begin{example}
  Let $ \ctb{El}$ be a type-level constant, and consider the rule \[
\ctb{El}~(\ctb{Prod}~A~B) \red \Pi x : \ctb{El}~A. \ctb{El}~(B~x)
  \] traditionally used to build \Dedukti{} encodings (as in \cite{dowek2007}). Note that here we write $ \alpha~\vec{l}$ for $ \alpha[\vec{l}] $, to ease the notation. We then have \[
\ctb{El}~(\ctb{Prod}~\ctb{Nat}~(\lambda x. \ctb{Nat})) \equiv \Pi x : \ctb{El}~\ctb{Nat}. \ctb{El}~((\lambda x. \ctb{Nat})~x) \equiv  \ctb{El}~\ctb{Nat} \to \ctb{El}~\ctb{Nat}  
\] but $ \tycont{\ctb{El}~(\ctb{Prod}~\ctb{Nat}~(\lambda x:.\ctb{Nat}))} = * $ and $ \tycont{\ctb{El}~\ctb{Nat} \to \ctb{El}~\ctb{Nat}}= * \to * $.
\end{example}

If we were to define the arity of a type\footnote{Note that this concept is different from the notion of arity of constants, as defined in Section \ref{sec:dedukti}.} as the number of consecutive arrows (that is, of $ \Pi $s), then we realize that the problem here is that rules such as $ \ctb{El}~(\ctb{Prod}~A~B) \red \Pi x : \ctb{El}~A. \ctb{El}~(B~x) $ do not preserve the arity. Indeed, $ \ctb{El}~(\ctb{Prod}~A~B) $ has arity $ 0 $ because it has no arrows, whereas $ \Pi x : \ctb{El}~A. \ctb{El}~(B~x) $ has arity $ 1 $ as it has one arrow\footnote{Using a different notation for the dependent product, we can write this type as $ (x : \ctb{El}~A) \to \ctb{El}~(B~x)$, which may help to clarify this assertion.}. As the left-hand side of a type-level rule always has arity $ 0 $ (because it is of the form $ \alpha[\vec{l}] $), to remove these unwanted cases we need for their right-hand sides to also have arity $ 0 $. This motivates the following definition.

\begin{definition}[Arity preserving]\label{arity-preserving}
  $ \mathscr{R} $ is said to be arity-preserving\footnote{More precisely, this definition also depends on the signature $ \Sigma $, as this is used to define which constants are type-level.} if, for every type-level rewrite rule in $  \mathscr{R} $, the right-hand side is in the following grammar, where $ \vec{M}, N, A $ are arbitrary. \[
 R ::= \alpha[\vec{M}]  \mid R~N \mid \lambda x: A. R    
\]
\end{definition}

It turns out that this definition, together with confluence of $ \beta\mathscr{R} $, will be enough to show that the translation preserves typing, and also non-termination. Therefore, throughout the rest of this section we suppose the following assumptions.

\begin{assumption}
  \label{assumption-sn-dedukti}
  $ \beta\mathscr{R} $ is confluent and $ \mathscr{R} $ is arity preserving.
\end{assumption}

Note that if $ \mathscr{R} $ is arity preserving then it is also well-formed, and thus we can in particular also use the stratification theorem.

\subsection{Proof of Strong Normalization of $ \beta $ in Dedukti}
\label{subsec:label}

We  start with the following key lemma, which ensures that convertible types are erased into the same simple type by $ \tycont{-} $.

\begin{lemma}[Key property]\label{key-property} ~
  \begin{enumerate}  
\item If $ \tycont{A} $ is defined, then for all $ N $, $ \tycont{A(N/x)} $ is also defined and $ \tycont{A}=\tycont{A(N/x)} $.
  \item If $ A \red A'$ and $ \tycont{A} $ is defined, then $ \tycont{A'} $ is also and $ \tycont{A}=\tycont{A'} $.
  \item If $ A \equiv A' $ and $ \tycont{A},\tycont{A'} $ are well defined, then $ \tycont{A}=\tycont{A'} $.
  \end{enumerate}
\end{lemma}
\begin{proof}
  \begin{enumerate}
  \item By induction on $ A $.
  \item By induction on the rewrite context. For the base case of $ \beta $, we use part 1. For the base case of a rule in $ \mathscr{R} $, this rule needs to be type-level, of the form $ \alpha[\vec{l}]\red r $. Note that for every substitution $ \sigma $, we have $ \tycont{\alpha[\vec{l} \{\sigma \}]} = *$. Thus, it suffices to show that for every $ \sigma $, $ \tycont{r \{ \sigma \}} $ is defined and equal to $ * $, which is done by induction on the grammar of Definition \ref{arity-preserving}.
  \item Follows from confluence and part 2.\qedhere
  \end{enumerate}
\end{proof}

With the key property in hand, we can show that the erasure preserves typing.

\begin{theorem}[Preservation of typing]
  \label{pres_typing_cont}
If $ \Sigma; \Gamma \vdash M : A $ and $ A \neq \Kind $, then there is $ \Sigma' \subseteq \Sigma $ such that  $ \Gamma_\pi,\tycont{\Sigma'},\tycont{\Gamma} \vdash_\lambda \cont{M} : \tycont{A} $
\end{theorem}
\begin{proof}
First note that for all $ x : A \in \Gamma $, we have $ \Sigma; \Gamma \vdash A : \Type $, thus by syntactic stratification $ A $ is a type-family, and thus $ \tycont{\Gamma} $ is well-defined.  We proceed by induction on the derivation. The base cases Var and Sort are trivial.

  \textbf{Cons}: We have

\begin{center}
\AxiomC{$\Sigma;\Gamma~\texttt{well-formed}$}
\AxiomC{$\Sigma;\Delta \vdash A : s $}
\AxiomC{$\Sigma;\Gamma \vdash \vec{M} : \Delta $}  
\LeftLabel{$c [\Delta] : A \in \Sigma$}
\RightLabel{\texttt{Cons}}
\TrinaryInfC{$\Sigma;\Gamma \vdash c[\vec{M}] : A \{\vec{M} \}  $}
\DisplayProof
\end{center}

We first show that $ \tycont{c[\Delta] : A} $ is defined. Write $ \Delta= x_1: A_1,...,x_n:A_n$. First note that $ \Sigma; \Delta \vdash A : s $ implies that for all $ x_i : A_i \in \Delta $ we have $ A_i : \Type $, hence by stratification each $ A_i $ is a type family and $ \tycont{-} $ is defined for all of them. Moreover, by stratification $  A : s $ implies that $ A $ is either a type family or kind, hence $ \tycont{A} $ is defined. Hence, $ \tycont{c[\Delta]:A} = c : \tycont{A_1} \to ... \tycont{A_n} \to \tycont{A} $ is well-defined.

We now proceed with the main proof obligation. By IH for $ i = 1,...,n $ we have $ \Sigma_i \subseteq \Sigma $ such that $ \Gamma_\pi , \tycont{\Sigma_i} , \tycont{\Gamma} \vdash \cont{M_i} : \tycont{A_i \{M_1/x_1 \}... \{M_{i-1}/x_{i-1} \}}$. Then, by Lemma \ref{key-property} we have $ \tycont{A_i \{M_1/x_1 \}... \{M_{i-1}/x_{i-1} \}} = \tycont{A_i} $. Therefore, by taking \[
 \Sigma' = c[\Delta] : A \cup \Sigma_1 \cup ... \cup\Sigma_n
\] we can derive $ \Gamma_\pi, \tycont{\Sigma'}, \tycont{\Gamma} \vdash_\lambda c~\cont{\vec{M}} : \tycont{A} $. Because $ \tycont{A} = \tycont{A \{ \vec{M}\}} $, the result follows.

  \textbf{Conv}: We have

  \begin{center} 
  \AxiomC{$\Sigma; \Gamma \vdash M : A $}
  \AxiomC{$\Sigma; \Gamma \vdash B : s $}
  \AxiomC{$A \equiv B $}    
\RightLabel{\texttt{Conv}}
\TrinaryInfC{$\Sigma;\Gamma \vdash M : B  $}
\DisplayProof
\end{center}
First note that $ A $ cannot be $ \Kind $. Indeed, by confluence we would have $ B \red^* \Kind $, but by syntactic stratification $ B $ is either a kind or a type family.  As kinds and type families are in particular pre-kinds and pre-type families, $ B $ is one those. But as they are closed under rewriting, this would imply that $ \Kind $ is a pre-kind or a pre-type family, absurd.

Therefore, by IH we have $ \Gamma_\pi,\tycont{\Sigma'},\tycont{\Gamma} \vdash_\lambda \cont{M} : \tycont{A} $ for some $ \Sigma' \subseteq \Sigma $. Moreover, by syntactic stratification $ A, B $ are kinds or type families, thus $ \tycont{-} $ is defined for them. By the Key Property  $ A \equiv B $ implies $ \tycont{A}=\tycont{B} $ and thus the result follows. 

  \textbf{Prod}: We have
  \begin{center} 
  \AxiomC{$\Sigma; \Gamma \vdash A : \Type $}
  \AxiomC{$\Sigma; \Gamma, x : A \vdash B : s $}  
\RightLabel{\texttt{Prod}}
\BinaryInfC{$\Sigma;\Gamma \vdash \Pi x : A . B : s  $}
\DisplayProof
\end{center}
If $ s = \Kind $ there is nothing to show, thus we consider $ s = \Type $. By IH, for some $ \Sigma', \Sigma'' \subseteq \Sigma  $ we have $ \Gamma_\pi,\tycont{\Sigma'},\tycont{\Gamma} \vdash_\lambda \cont{A}: * $ and $ \Gamma_\pi,\tycont{\Sigma''},\tycont{\Gamma}, x : \tycont{A} \vdash_\lambda \cont{B}: * $. We thus get $\Gamma_\pi,\tycont{\Sigma''},\tycont{\Gamma} \vdash_\lambda \lambda x.\cont{B}: \tycont{A} \to * $, and therefore $ \Gamma_\pi,\tycont{\Sigma'\cup\Sigma''},\tycont{\Gamma}\vdash_\lambda \pi_{\tycont{A}}~\cont{A}~(\lambda x.\cont{B}): * $

  \textbf{Abs}: We have
\begin{center}
  \AxiomC{$\Sigma; \Gamma \vdash A : \Type $}
  \AxiomC{$\Sigma; \Gamma, x : A \vdash B : s $}  
  \AxiomC{$\Sigma; \Gamma, x : A \vdash M : B $}
  \RightLabel{\texttt{Abs}}
\TrinaryInfC{$\Sigma;\Gamma \vdash \lambda x : A . M :\Pi x : A . B  $}
\DisplayProof
\end{center}
By IH,  for some $ \Sigma', \Sigma'' \subseteq \Sigma $ we have $ \Gamma_\pi,\tycont{\Sigma'},\tycont{\Gamma} \vdash_\lambda \cont{A}: *$ and $ \Gamma_\pi,\tycont{\Sigma''},\tycont{\Gamma}, x : \tycont{A} \vdash_\lambda \cont{M}: \tycont{B}$, from which we deduce $ \Gamma_\pi,\tycont{\Sigma''},\tycont{\Gamma} \vdash_\lambda \lambda x.\cont{M}: \tycont{A} \to \tycont{B}$. By adding some spurious variable $ z $ of type $ * $ to the context and abstracting over it, we get $ \Gamma_\pi,\tycont{\Sigma''},\tycont{\Gamma} \vdash_\lambda \lambda z.\lambda x.\cont{M}: * \to \tycont{A} \to \tycont{B}$. Finally, we use application to conclude $ \Gamma_\pi,\tycont{\Sigma'\cup\Sigma''},\tycont{\Gamma} \vdash_\lambda (\lambda z.\lambda x.\cont{M})\cont{A}:\tycont{A} \to \tycont{B}$.

  \textbf{App}: We have
\begin{center}
  \AxiomC{$\Sigma;\Gamma \vdash M : \Pi x : A . B  $}
  \AxiomC{$\Sigma; \Gamma \vdash N : A $}
  \RightLabel{\texttt{App}}
\BinaryInfC{$\Sigma;\Gamma \vdash M N : B \{N/x\} $}
\DisplayProof
\end{center}
By IH, for some $ \Sigma', \Sigma'' \subseteq \Sigma $ we deduce $ \Gamma_\pi,\tycont{\Sigma'},\tycont{\Gamma} \vdash_\lambda \cont{M}: \tycont{A} \to \tycont{B} $ and $ \Gamma_\pi,\tycont{\Sigma''},\tycont{\Gamma} \vdash_\lambda \cont{N}: \tycont{A} $. By application, we get $\Gamma_\pi,\tycont{\Sigma'\cup\Sigma''},\tycont{\Gamma} \vdash_\lambda \cont{N}\cont{M}: \tycont{B} $, and as $ \tycont{B} = \tycont{B \{N/x \}} $ the result follows.
\end{proof}

\begin{proposition}[Preservation of non-termination]  \label{pres_beta_cont}
  Let $ M $ be an object or type family. 
\begin{enumerate}
\item If $ N $ is an object, then $ M \{N/x \} $ is an object or type family and $ \cont{M \{N/x \}} = \cont{M} \{\cont{N}/x\} $.
\item If $ M \red_\beta N $ then $ N $ is an object or type family and $ \cont{M} \red_\beta^+ \cont{N} $.
\end{enumerate}
\end{proposition}
\begin{proof}
  \begin{enumerate}
  \item By induction on $M$, and using Lemma \ref{key-property} for the case $ M = \Pi x : A. B $.
  \item By induction on the rewriting context. For the base case, we have $ M = (\lambda x : A. M_1)M_2 \red M_1 \{M_2/x\}$ and thus $ \cont{M} = (\lambda z. \lambda x. \cont{M_1})\cont{A}\cont{M_2} $. As $ z $ is not free in $ \cont{M_1} $, we have $\cont{M} = (\lambda z. \lambda x. \cont{M_1})\cont{A}\cont{M_2} \red (\lambda x. \cont{M_1}) \cont{M_2} \red \cont{M_1}\{\cont{M_2}/x\} $. By part 1, $ \cont{M_1 \{M_2 /x \} } $ is well-defined and equal to $ \cont{M_1}\{\cont{M_2}/x\} $.

    The induction steps are all similar, we present two of them to show the idea. If $ M=\lambda x : A. M' \red \lambda x : A'. M' = N $, where $ A \red A' $, then by IH we have $ \cont{A} \red^+ \cont{A'} $, and thus $ \cont{M}=(\lambda z. \lambda x. \cont{M'})\cont{A} \red^+ (\lambda z. \lambda x. \cont{M'})\cont{A'} = \cont{N}$. If $ M=\Pi x: A. B \red \Pi x : A'. B = N $, where $ A \red A' $, then by IH we have $ \cont{A} \red^+ \cont{A'} $. By the key property, this implies $ \tycont{A}=\tycont{A'} $, and thus $ \cont{M}=\pi_{\tycont{A}}~\cont{A}~(\lambda x.\cont{B}) \red^+ \pi_{\tycont{A'}}~\cont{A'}~(\lambda x.\cont{B})$.
\qedhere
  \end{enumerate} 
\end{proof}

\begin{theorem}[$ \beta $ is SN in \Dedukti]
  \label{norm_beta_dk}If $ \beta\mathscr{R} $ is confluent and $ \mathscr{R} $ is arity-preserving, then $ \beta $ is strongly normalizing for well-typed terms in \Dedukti.
\end{theorem}
\begin{proof}
  Suppose that $ M $ satisfies $ \Sigma; \Gamma \vdashdk M : A $ and there is an infinite sequence $M = M_1 \red_\beta M_2 \red_\beta  M_3 \red_\beta ...$ starting from $ M $. We now show that for some $ N $, $ \cont{N} $ is well-typed in the simply-typed $ \lambda $-calculus and an infinite sequence starts from $ \cont{N} $.

  If $ A \neq \Kind $, then this follows directly from Proposition \ref{pres_typing_cont} by taking $ N=M $. If $ A = \Kind $, then $ M  $ is of the form  $ \Pi \vec{x} : \vec{B}. \Type $, and as there are finitely many $ B $s and they are all type families, we conclude that there is a type family $ B_i $ from which an infinite sequence starts. We can thus take $ N = B_i $ and apply Proposition \ref{pres_typing_cont} to get the result.

  Now note that as objects and type-families are closed under $ \beta $, then $ \cont{-} $ is defined for all elements in the sequence. Therefore, by taking the image of this infinite sequence under $ \cont{-} $ we also get an infinite sequence, by Proposition \ref{pres_beta_cont}. This is a contradiction with the strong normalization of $ \beta $ in the simply typed $ \lambda $-calculus, hence the result follows.
\end{proof}

}

\section{Pure Type Systems}
\label{sec:pts}

Pure type systems (or PTSs) is a class of type systems that generalizes many other systems, such as the Calculus of Constructions and System F. They are parameterized by a set of \textit{sorts} $ \mathcal{S} $ (refereed to by the letter $ s $) and two relations $ \mathcal{A}\subseteq\mathcal{S}^2, \mathcal{R}\subseteq\mathcal{S}^3 $. In this work we restrict ourselves to functional PTSs, for which $ \mathcal{A} $ and $ \mathcal{R} $ are functional relations. This restriction covers almost all of PTSs used in practice, and gives a much more well behaved metatheory.

In this paper we consider a variant of PTSs with explicit parameters. That is, just like when taking the projection of a pair $\pi^1(p) $ we can make explicit all parameters and write $ \pi^1(A, B, p) $ where $ p : A \times B $, we can also write $ \lambda(A,\bind{x}B,\bind{x}M) $ instead of $ \lambda x : A. M $ and $ @(A,\bind{x}B,M,N) $ instead of $ M N $. Moreover, if $ (-)\times(-) $ is a universe-polymorphic definition, we should also  write $ \pi^1_{s_A,s_B}(A,B,p) $ to make explicit the sort parameters. As in PTSs the dependent product is used across multiple sorts, we then should also write $ \lambda_{s_A,s_B}(A,\bind{x}B,\bind{x}M) $, $ @_{s_A,s_B}(A,\bind{x}B,M,N) $ and $ \Pi_{s_A,s_B}(A,\bind{x}B) $. To be more direct, we render explicit the parameters on the dependent product type and on its constructor (abstraction) and eliminator (application). Because of this interpretation in which we are rendering the parameters of $ \lambda $ and $ @ $ explicit, we name this version of PTSs as Explicitly-typed Pure Type Systems (EPTSs).

Reduction is then defined  by the context closure of the $ \beta $ rules\footnote{We consider a linearized variant of the expected non-left linear rule $ @_{s_1,s_2}( A, \bind{x} B, \lambda_{s_1,s_2}( A, \bind{x} B, \bind{x} M),  N) \red M \{ N/x \} $, which is non-confluent in untyped terms. By linearizing it, we get a much more well-behaved rewriting system, where confluence holds for all terms. Moreover, whenever the left hand side is well-typed, the typing constraints impose $ A \equiv A' $ and $ B \equiv B' $.} \[
@_{s_1,s_2}( A, \bind{x} B, \lambda_{s_1,s_2}( A', \bind{x} B', \bind{x} M),  N) \red M\{N/x\}
  \] given for each $ (s_1,s_2,s_3) \in \mathcal{R} $. Typing is given by the rules in Figure \ref{typing-epts}.

\begin{figure}[H]
{\small
\begin{center}
  \AxiomC{}
\RightLabel{\textsc{Empty}}
\UnaryInfC{$-~\WF$}
\DisplayProof
\hskip 3em
\AxiomC{$\Gamma \vdash A : s$}
\RightLabel{\textsc{Decl}}
\LeftLabel{$x \notin \Gamma$}
\UnaryInfC{$\Gamma, x : A ~\WF $}
\DisplayProof
\end{center}
\begin{center}
  \AxiomC{$\Gamma \vdash M : A $}
  \AxiomC{$ \Gamma \vdash B : s $}  
  \LeftLabel{$A \equiv B$}
  \RightLabel{\textsc{Conv}}
\BinaryInfC{$\Gamma \vdash M : B $}
\DisplayProof
\hskip 3em
  \AxiomC{$\Gamma $ well-formed}
  \LeftLabel{$(s_1, s_2) \in \mathcal{A} $}
  \RightLabel{\textsc{Sort}}
\UnaryInfC{$\Gamma \vdash s_1 : s_2$}
\DisplayProof
\end{center}
\begin{center}
\AxiomC{$\Gamma~\WF$}
\LeftLabel{$ x : A  \in \Gamma $}
\RightLabel{\textsc{Var}}
\UnaryInfC{$\Gamma \vdash x : A  $}
\DisplayProof
\hskip 3em
  \AxiomC{$\Gamma \vdash A : s_1 $}
  \AxiomC{$\Gamma, x : A \vdash B : s_2 $}
  \LeftLabel{$(s_1, s_2, s_3) \in \mathcal{R} $}
\RightLabel{\textsc{Prod}}
\BinaryInfC{$\Gamma \vdash \Pi_{s_1,s_2}(A, \bind{x} B) : s_3 $}
\DisplayProof
\end{center}
\begin{center}
  \AxiomC{$\Gamma \vdash A : s_1 $}
  \AxiomC{$\Gamma ,x : A \vdash B :s_2 $}  
  \AxiomC{$\Gamma , x : A \vdash M : B $}
  \LeftLabel{$(s_1, s_2, s_3) \in \mathcal{R} $}  
  \RightLabel{\textsc{Abs}}
\TrinaryInfC{$\Gamma \vdash \lambda_{s_1,s_2} (A, \bind{x} B, \bind{x} M) : \Pi_{s_1,s_2}(A, \bind{x} B) $}
\DisplayProof
\end{center}
\begin{center}
  \AxiomC{$\Gamma \vdash A : s_1 $}
  \AxiomC{$\Gamma ,x : A \vdash B :s_2 $}
  \AxiomC{$\Gamma \vdash N : A $}
  \AxiomC{$\Gamma \vdash M :\Pi_{s_1,s_2}(A, \bind{x} B)$}
  \LeftLabel{$(s_1, s_2, s_3) \in \mathcal{R} $}    
    \RightLabel{\textsc{App}}
\QuaternaryInfC{$\Gamma \vdash @_{s_1,s_2}(A, \bind{x} B,  M,  N) : B \{ N/x \} $}
\DisplayProof
\end{center}}
\caption{Typing rules for Explicitly-typed Pure Type Systems}
\label{typing-epts}
\end{figure}


This modification is just a technical change that will help us during the translation, as our encoding  needs the data of such parameters often left implicit. Other works such as \cite{mellies} and \cite{siles_herbelin} also consider similar variants, though none of them corresponds exactly to ours. Therefore,  we had to develop the basic metatheory of our version in \cite{epts}, and we have found that  the usual meta-theoretical properties of functional PTSs  are preserved when moving to the explicitly-typed version\toshort{ (see Appendix \ref{sec:meta_epts})}. More importantly, by a proof that uses ideas present in \cite{siles_herbelin}, we have shown the following equivalence.

Let $ |-| $ be the erasure map defined in the most natural way from an EPTS to its corresponding PTS. Moreover, for a system $ X $ let $\Lambda(\Gamma \vdash_X \_ : A)$ be  the set of $ M \in \Lambda_X $ with $ \Gamma \vdash_{X} M : A  $. Finally, let $ \equiv_I $ be defined by  $ M \equiv_I N $ iff $ |M| = |N|  $ and $ M \equiv N $.

\begin{theorem}[Equivalence between PTSs and EPTSs\cite{epts}]
  \label{equivalence_PTS_EPTS}
Consider a functional PTS. If $ \Gamma \vdash_{PTS} A~type $, then there are $ \Gamma', A' $ with $ |\Gamma'| = \Gamma, |A'| = A $ such that we have a bijection \[
\Lambda(\Gamma \vdash_{PTS} \_ : A) \simeq \Lambda(\Gamma' \vdash_{EPTS} \_ : A')/ \equiv_I
\]
\end{theorem}

\tolong{We note that functional EPTSs satisfy the following basic properties, whose proofs can be found in \cite{epts}.

\begin{proposition}[Weakening]
  \label{weakening}
  Let $ \Gamma \sqsubseteq \Gamma' $ with $ \Gamma'~\WF $. If $ \Gamma \vdash M : A $ then $ \Gamma' \vdash M : A $.
\end{proposition}

\begin{proposition}[Inversion]
  \label{inversion}
  If $ \Gamma \vdash M : C $ then
  \begin{itemize}
  \item If $ M = x $, then
    \begin{itemize}
    \item $ \Gamma~\textup{well-formed}$ with a smaller derivation tree
    \item there is $x$ with $ x : A \in \Gamma $ and $ C \equiv A $
    \end{itemize}
  \item If $ M = s $, then there is $s'$ with $ (s,s') \in \mathcal{A} $ and $ C \equiv s' $
  \item If $ M = \Pi_{s_1,s_2}(A,\bind{x}B) $ then
    \begin{itemize}
    \item $ \Gamma \vdash A:s_1 $ with a smaller derivation tree
    \item $ \Gamma,x:A \vdash B : s_2 $ with a smaller derivation tree
    \item there is $ s_3 $ with $ (s_1,s_2,s_3)\in \mathcal{R} $ and $ C \equiv s_3 $
    \end{itemize}
  \item If $ M = \lambda_{s_1,s_2}(A,\bind{x}B,\bind{x}N) $ then
    \begin{itemize}
    \item $ \Gamma \vdash A : s_1 $ with a smaller derivation tree
    \item $ \Gamma, x : A \vdash B : s_2 $ with a smaller derivation tree
    \item there is $ s_3 $ with $ (s_1,s_2,s_3) \in \mathcal{R} $      
    \item $ \Gamma, x:A \vdash N:B $ with a smaller derivation tree
    \item $C \equiv \Pi_{s_1,s_2}(A,\bind{x}B) $
    \end{itemize}
  \item If $ M = @_{s_1,s_2}(A, \bind{x} B,  N_1,  N_2) $ then
    \begin{itemize}
    \item $ \Gamma \vdash A : s_1 $ with a smaller derivation tree
    \item $ \Gamma, x : A \vdash B : s_2 $ with a smaller derivation tree
    \item there is $ s_3 $ with $ (s_1,s_2,s_3) \in \mathcal{R} $      
    \item $ \Gamma \vdash N_1 : A $ with a smaller derivation tree
    \item $ \Gamma \vdash N_2 : \Pi_{s_1,s_2}(A,\bind{x}B) $ with a smaller derivation tree
    \item $ C \equiv B\{N_2/x\} $
    \end{itemize}
  \end{itemize}
\end{proposition}

\begin{proposition}[Uniqueness of types]
  \label{uniqueness_of_types}
If $ \Gamma \vdash M : A $ and  $ \Gamma \vdash M : B  $ we have $ A \equiv B $.
\end{proposition}

\begin{corollary}[Uniqueness of sorts]
  \label{uniqueness_of_sorts}
If $ \Gamma \vdash M : s $ and  $ \Gamma \vdash M : s'  $ we have $ s=s' $.
\end{corollary}

\begin{proposition}[Conv in context]
  \label{conv_in_context}
  Let $ A \equiv A' $ and $ \Gamma \vdash A' : s $. We have
  \begin{itemize}
  \item $ \Gamma,x:A,\Gamma'~\textup{well-formed} \Rightarrow \Gamma,x:A',\Gamma'~\textup{well-formed} $
  \item $ \Gamma,x:A,\Gamma'\vdash M : B \Rightarrow \Gamma,x:A',\Gamma' \vdash M :B $
  \end{itemize}  
\end{proposition}

\begin{proposition}[Substitution in judgment]
  \label{substitution_lemma}
  Let $ \Gamma \vdash N : A $. We have
  \begin{itemize}
  \item $ \Gamma,x:A,\Gamma'~\textup{well-formed} \Rightarrow \Gamma,\Gamma'\{N/x\}~\textup{well-formed} $
  \item $ \Gamma,x:A,\Gamma'\vdash M : B \Rightarrow \Gamma,\Gamma'\{N/x\} \vdash M \{N/x\} : B\{N/x\} $
  \end{itemize}
\end{proposition}

}

\section{Encoding EPTSs in Dedukti}
\label{sec:encoding}

This section presents our encoding of functional EPTSs in \Dedukti. In order to ease the notation, from now one we write $ c~\vec{M} $ for $ c[\vec{M}]$. The basis for the encoding is given by a theory $ (\Sigmaepts,\Repts) $ which we will construct step by step here.

Pure Type Systems (explicitly-typed or not) feature two kinds of types: dependent products and universes. We start by building the representation of the latter. For each $ s \in \mathcal{S} $ we declare a type $ \U{s} $ to represent the type of elements of $ s $. However, as the terms $ A $ with $ \Gamma \vdash_{EPTS} A : s $ are themselves types, we also need to declare a function $ \El{s} $ which maps each such $ A $ to its corresponding type. As for each $ (s_1, s_2)  \in \mathcal{A}$ we have $ \vdash_{EPTS} s_1 : s_2 $, we also  declare a constant  $ \uu{s_1} $ in $ \U{s_2} $ to represent this. Finally, as the sorts $ s_1 $ with $ (s_1, s_2) \in \mathcal{A} $ now can be  represented by both $ \U{s_1} $ and $ \El{s_2}~\uu{s_1} $, we add a rewrite rule to identify these representations. This encoding resembles the  definition of universes in type theories \textit{à la} Tarski, and also follows traditional representations of universes in \Dedukti{} as in \cite{dowek2007}.

\noindent\parbox{.5\textwidth}{
  \dkbox{55}{
  &\ctb{U_s} : \Type \\
  &\ctb{El_s}[A: \ctb{U_s}] : \Type & \text{for $ s \in \mathcal{S} $}    
}}
\parbox{.5\textwidth}{
  \dkbox{55}{
    &\uu{s_1} : \ctb{U_{s_2}} \\
    &\ctb{El_{s_2}}~\uu{s_1} \red_{\uu{s_1}\text{-red}}    \U{s_1} & \text{for $ (s_1,s_2) \in \mathcal{A} $} 
}}

We now move to the representation of the dependent product type. We first declare a constant to represent the type formation rule for the dependent product. 

\dkbox{130}{  &\ctb{Prod_{s_1,s_2}}[A : \U{s_1};  B : \ctb{El_{s_1}}~A \to \U{s_2}] : \U{s_3} & \text{for $ (s_1,s_2,s_3) \in \mathcal{R} $}}

Traditional \Dedukti~encodings would normally continue here by introducing the rule $ \El{s_3}~(\Prod{s_1}{s_2}~A~B) \red \Pi x : \El{s_1}~A. \El{s_2}~(B~x) $, identifying the dependent product of the encoded theory with the one of \Dedukti, thus allowing for the use of the framework's abstraction, application and $ \beta $ to represent the ones of the encoded system. We instead keep them separate and declare constants representing the introduction and elimination rules for the dependent product being encoded, that is, representing abstraction and application. 

\noindent\dkbox{130}{
    &\abs{s_1}{s_2}[A : \U{s_1} ; B : \El{s_1}~A \to \U{s_2};M:\Pi x : \El{s_1}~A. \El{s_2}~(B~x)] : \El{s_3}(\Prod{s_1}{s_2}~A~B) \\
  &\app{s_1}{s_2}[A : \U{s_1}; B : \El{s_1}~A \to \U{s_2}; M : \El{s_3}(\Prod{s_1}{s_2}~A~B); N : \El{s_1}~A] : \El{s_2}(B~N)\\
  &\app{s_1}{s_2}~A~B~(\abs{s_1}{s_2}~A'~B'~M)~N \red_{\betarule{s_1}{s_2}} M~N  \hspace{9.7em}\text{for $ (s_1,s_2,s_3) \in \mathcal{R}$}
}

We note that this idea is also hinted in \cite{assaf}, though they did not pursue it further. This approach also resembles the one of the Edinburgh Logical Framework (\ELF) \cite{ELF} in which the framework's abstraction is used exclusively for binding. We are however able to encode computation directly as computation with the rule $ \betarule{s_1}{s_2} $, whereas the \ELF{}  handles computation by encoding it as an equality judgment, thus introducing explicit coercions in the terms. Some other  variants such as \cite{harper2021equational} prevent the introduction of such coercions, but computation is still represented by an equality judgment instead of being represented by computation.

This finishes the definition of the theory $ (\Sigmaepts,\Repts) $. Now we ready to define the translation function $ \trans{-} $.
{\small\begin{align*}
  &\trans{ x } = x\\
  &\trans{ s } = \uu{s} \\
  &\trans{ \Pi_{s_1,s_2}(A, \bind{x} B) } = \ctb{Prod_{s_1,s_2}}~\trans{ A }~(\lambda x : \El{s_1}~\trans{A}. \trans{ B })\\
  &\trans{ \lambda_{s_1,s_2}( A, \bind{x} B, \bind{x} M) } = \ctb{abs_{s_1,s_2}}~\trans{ A }~(\lambda x : \El{s_1}~\trans{A}. \trans{ B })~(\lambda x : \El{s_1}~\trans{A}. \trans{ M })\\
  &\trans{ @_{s_1,s_2}( A, \bind{x} B,  M,  N) } = \ctb{app_{s_1,s_2}}~\trans{ A }~(\lambda x : \El{s_1}~\trans{A}. \trans{ B })~\trans{ M }~\trans{ N }
\end{align*}}%

We also extend $ \trans{-} $ to well-formed contexts by the following definition. Note that because we are dealing with functional EPTSs, the sort of $ A $ in $ \Gamma $ is unique, hence the following definition makes sense.
{\small\begin{align*}
         &\trans{-} = -\\
         &\trans{\Gamma, x : A} = \trans{\Gamma}, x : \El{s_A}~\trans{A}\quad\text{where } \Gamma \vdash A : s_A
\end{align*}}%

\begin{remark}
  Note that in the definition of $ \trans{-} $ it was essential for $ \lambda $ and $ @ $ to make explicit the types $ A $ and $ B $, as the constants $ \abs{s_1}{s_2} $ and $ \app{s_1}{s_2} $ require their translations. Had we had for instance just $ \lambda x : A. M $, we could then make the translation dependent on $ \Gamma $ and take a $ B $ such that $ \Gamma, x : A\vdash M : B $. However, because $ \trans{-} $ is defined by induction and  $ B $ is not a subterm of $ \lambda x : A. M $, we cannot apply $ \trans{-} $ to $ B $. Therefore,  when doing an encoding in \Dedukti{}  one should  first render explicit the needed data before translating, and then show an equivalence theorem between the explicit and implicit versions (in our case, Theorem \ref{equivalence_PTS_EPTS}).

  Moreover, note that by also making the sorts explicit in $ \lambda_{s_1,s_2}, @_{s_1,s_2}, \Pi_{s_1,s_2} $ our translation can be defined purely syntactically. If this information were not in the syntax, we  could still define  $ \trans{-} $ by making it dependent on $ \Gamma $, as is usually done with traditional \Dedukti{} encodings\cite{dowek2007}. Nevertheless, this complicates many proofs, as each time we apply $ \trans{-}_\Gamma $ to a term we need to know it is well-typed in $ \Gamma $. Moreover, properties which should concern all untyped terms (such as preservation of computation) would then be true only for well-typed ones.
\end{remark}

In order to understand more intuitively how the encoding works, let's look at an example. 

\begin{example}
  Recall that System F can be defined by the sort specification $ \mathcal{S} = \{Type,Kind\}, \mathcal{A} = \{(Type, Kind)\}, \mathcal{R} = \{(Type,Type, Type), (Kind,Type,Type)\}$. In this EPTS, we can express the polymorphic identity function, traditionally written as $ \lambda A: Type. \lambda x : A.x $, by \[
\lambda_{Kind,Type}(Type,\bind{A}\Pi_{Type,Type}(A,\bind{x}A), \bind{A} \lambda_{Type, Type}(A, \bind{x}A, \bind{x}x))
\]This term is represented in our encoding by \[
\abs{Kind}{Type}~\uu{Type}~(\lambda A.\Prod{Type}{Type}~A~(\lambda x. A))~(\lambda A.\abs{Type}{Type}~A~(\lambda x. A)~(\lambda x . x))
\]where we omit the type annotations in the abstractions, to improve readability. 
\end{example}

\section{Soundness}
\label{sec:soundness}

An encoding is said to be sound when it preserves the typing relation of the original system. In this section we will see that our encoding has this fundamental property. We start by establishing some conventions in order to ease notations.

\begin{convention}
  We establish the following notations.
  \begin{itemize}
  \item We write $ \Sigma;\Gamma \vdashdk M : A $ for a \Dedukti{} judgment and $ \Gamma \vdash M : A $  for an EPTS judgment 
  \item As the same signature $ \Sigmaepts $ is used everywhere, when referring to $ \Sigmaepts;\Gamma \vdashdk M : A $ we omit it and write $ \Gamma \vdashdk M : A $.
  \end{itemize}
\end{convention}

Before showing soundness, we start by establishing some  basic results. 

\begin{proposition}[Basic properties]\label{basic-properties}
  We have the following basic properties.
  \begin{enumerate}
  \item Confluence: The rewriting rules of the encoding are confluent with $ \beta $.
  \item Well-formedness of the signature: For all $ c[\Delta] : A \in \Sigmaepts $, we have $ \Delta\vdashdk A : s $.
  \item Subject reduction for $ \beta $: If $ \Gamma \vdashdk M : A $ and $ M \red_\beta M' $ then $ \Gamma \vdashdk M' : A $.
  \item Strong normalization for $ \beta $: If $ \Gamma \vdashdk M : A $, the $ \beta $  is strongly normalizing for $ M $.
  \item Compositionality: For all $ M, N \in \Lambda_{EPTS}$ we have $ \trans{M} \{ \trans{N}/x \} = \trans{M \{ N/x \}} $.
  \end{enumerate}
\end{proposition}
\begin{proof}
\begin{enumerate}
\item The considered rewrite rules form an orthogonal combinatory reduction system, and therefore are confluent\cite{CRS}.
\item Can be shown for instance with \textsc{Lambdapi}\cite{lambdapi}, an implementation of \Dedukti.
\item Subject reduction of $ \beta $ is implied by  confluence of $ \beta\Repts $\cite{frederic-phd}.
\item $ \Repts $ is arity preserving and $ \beta\Repts $ is confluent, thus \thref{norm_beta_dk} applies.
\item By  induction on $ M $.\qedhere
\end{enumerate}
\end{proof}

\begin{remark}
We could also show subject reduction of our encoding, either using the method in \cite{subject_reduction} or \textsc{Lambdapi}\cite{lambdapi}. However, we will see that our proof does not actually require subject reduction of $ \Repts $. Therefore, we conjecture that our proof method can also be adapted to systems that do not satisfy subject reduction.
\end{remark}

\begin{lemma}[Preservation of computation]\label{preservation_computation}
  Let $ M, N \in \Lambda_{EPTS} $. We have
  \begin{enumerate}
  \item   $M \red N$ implies $\trans{ M } \red^* \trans{ N }$
  \item $M \equiv N$  implies $\trans{ M } \equiv \trans{ N }$
  \end{enumerate}
\end{lemma}
\begin{proof}
  Intuitively, the first part holds because a $ \beta $ step in the source system is represented by a $ \ctb{beta} $ step followed by a $ \beta  $ step in \Dedukti.  It is shown by induction on the rewriting context, using compositionality of $ \trans{-} $ for the base case. The second part follows by induction on $ \equiv $ and uses part 1.
\end{proof}

Recall that a sort $ s \in \mathcal{S} $ is said to be a top-sort if there is no $ s' $ with $ (s,s') \in \mathcal{A} $. The following auxiliary lemma allows us to switch between sort representations and is heavily used in the proof of soundness.

\begin{lemma}[Equivalence for sort representations]\label{equiv_sort}
If $ s $ is not a top-sort, then \[
\Gamma \vdashdk M : \U{s} \iff  \Gamma \vdashdk M : \El{s'}~\uu{s} 
\]where $ (s,s') \in \mathcal{A} $.
\end{lemma}

With all these results in hand, we can now show the soundness of our encoding.

\begin{theorem}[Soundness]
  \label{soundness}
  Let $ \Gamma $ be a context and $ M, A $ terms in an EPTS. We have
  \begin{itemize}
  \item If $ \Gamma~\WF $ then  $ \trans{\Gamma}~\textup{\texttt{well-formed}} $
  \item If $  \Gamma \vdash M : A  $ then \begin{itemize}
    \item if $ A  $ is a top-sort then $ \trans{\Gamma} \vdashdk \trans{ M } : \U{A}$
    \item else $ \trans{\Gamma} \vdashdk \trans{ M } : \El{s_A}~\trans{ A }$, where $ \Gamma \vdash A : s_A $
    \end{itemize}
    
  \end{itemize}
\end{theorem}
\begin{proof} 
  By structural induction on the proof of the judgment. \toshort{We present here only the case \textsc{Prod} to show the idea, the other cases are detailed in the long version in \cite{adequate}.}\tolong{Easy for the cases \textsc{Empty} and \textsc{Var}.}

  \tolong{\textbf{Case Decl}:  The proof ends with
  {\small\begin{center} 
  \AxiomC{$\Gamma \vdash A : s $}
  \LeftLabel{$x \notin A$}
\RightLabel{\textsc{Decl}}
\UnaryInfC{$\Gamma \vdash x : A$}
\DisplayProof
\end{center}}
From the IH we can derive  $ \trans{\Gamma} \vdashdk \El{s}~\trans{A} : \Type $, therefore we can apply \texttt{Decl} to get $ \trans{\Gamma}, x : \El{s}~\trans{A}~\texttt{well-formed} $. 

\textbf{Case Sort}: The proof ends with
  {\small\begin{center} 
      \AxiomC{$\Gamma~\WF $}
        \LeftLabel{$(s_1,s_2) \in \mathcal{A} $}
\RightLabel{\textsc{Sort}}
\UnaryInfC{$\Gamma \vdash s_1 : s_2$}
\DisplayProof
\end{center}}

By IH we have $ \trans{\Gamma}~\texttt{well-formed} $, therefore we can show $ \trans{\Gamma} \vdashdk \uu{s_1} : \U{s_2} $ using \texttt{Cons}. If $ s_2 $ is not a top-sort, we use Lemma \ref{equiv_sort} to show $ \trans{\Gamma} \vdashdk \uu{s_1} : \El{s_3}~\uu{s_2} $, where $ (s_2,s_3) \in \mathcal{A} $.

} \textbf{Case Prod}: The proof ends with
  {\small\begin{center} 
  \AxiomC{$\Gamma \vdash A : s_1 $}
  \AxiomC{$\Gamma, x : A \vdash B : s_2 $}
  \LeftLabel{$(s_1, s_2, s_3) \in \mathcal{R} $}
\RightLabel{\textsc{Prod}}
\BinaryInfC{$\Gamma \vdash \Pi_{s_1,s_2}(A, \bind{x} B) : s_3 $}
\DisplayProof
\end{center}}
By the IH and Lemma \ref{equiv_sort}, we have $\trans{\Gamma} \vdashdk \trans{ A } : \U{s_{1}}$ and $\trans{\Gamma}, x : \El{s_1}~\trans{A} \vdashdk \trans{ B } : \U{s_{2}}$. By \texttt{Abs} we get $\trans{\Gamma}\vdashdk \lambda x : \El{s_1}~\trans{A}. \trans{ B } : \El{s_1}~\trans{A} \to \U{s_{2}}$, therefore it suffices to apply \texttt{Cons} with $ \Prod{s_1}{s_2} $ to conclude \[
\trans{\Gamma} \vdashdk \Prod{s_1}{s_2}~\trans{A}~(\lambda x : \El{s_1}~\trans{A } . \trans{ B }) : \U{s_3}
\]If $ s_3 $ is not a top-sort, we then apply Lemma \ref{equiv_sort}. \toshort{\qedhere}

\tolong{\textbf{Case App:} The proof ends with
{\small\begin{center}
    \AxiomC{$\Gamma \vdash A : s_1 $}
    \AxiomC{$\Gamma ,x : A \vdash B :s_2 $}
    \AxiomC{$\Gamma \vdash M :\Pi_{s_1,s_2}(A, \bind{x} B)$}
    \AxiomC{$\Gamma \vdash N : A $}
    \LeftLabel{$(s_1, s_2, s_3) \in \mathcal{R} $}      
    \RightLabel{\textsc{App}}
    \QuaternaryInfC{$\Gamma \vdash @_{s_1,s_2}(A, \bind{x} B,  M,  N) : B(N/x) $}
    \DisplayProof
  \end{center}}
By the IH and Lemma \ref{equiv_sort}, we have $\trans{\Gamma} \vdashdk \trans{ A } : \U{s_{1}}$, $\trans{\Gamma}, x : \El{s_1}~\trans{A } \vdashdk \trans{ B } : \U{s_{2}}$, $\trans{\Gamma} \vdashdk \trans{ M } : \El{s_3}~(\Prod{s_{1}}{s_{2}}~\trans{ A }~(\lambda x : \El{s_1}~\trans{A } . \trans{ B }))$ and $\trans{\Gamma} \vdashdk \trans{ N } : \El{s_1}~\trans{ A } $. By \texttt{Abs} we get $\trans{\Gamma}\vdashdk \lambda x : \El{s_1}~\trans{A}. \trans{ B } : \El{s_1}~\trans{A} \to \U{s_{2}}$, therefore we can  apply \texttt{Cons} with $ \app{s_1}{s_2} $ to get \[
\trans{\Gamma} \vdashdk \app{s_1}{s_2}~\trans{A}~(\lambda x : \El{s_1}~\trans{A } . \trans{ B })~\trans{M}~\trans{N} : \El{s_2}~((\lambda x : \El{s_1}~\trans{A }. \trans{ B })~\trans{N})
\]Therefore, from \thref{reduce_type} with $ (\lambda x : \El{s_1}~\trans{A } . \trans{ B })\trans{N} \red  \trans{ B } \{ \trans{N} /x\}$ and compositionality of $ \trans{-} $ we get \[
\trans{\Gamma} \vdashdk \app{s_1}{s_2}~\trans{A}~(\lambda x : \El{s_1}~\trans{A } . \trans{ B })~\trans{M}~\trans{N} : \El{s_2}~\trans{ B \{N/x\}}
  \,.\]Finally, note that $ \Gamma \vdash N : A $ and $ \Gamma, x: A \vdash B : s_2 $ imply $ \Gamma \vdash B  \{N/x\} : s_2 $, thus $ B\{N/x\} $ is not a top-sort.

\textbf{Case Conv}: The derivation ends with
  {\small\begin{center}
  \AxiomC{$\Gamma \vdash M : B $}
  \AxiomC{$ \Gamma \vdash A : s $}  
  \LeftLabel{$A \equiv B$}
  \RightLabel{\textsc{Conv}}
\BinaryInfC{$\Gamma \vdash M : A $}
\DisplayProof
\end{center}}%
First note that by confluence and subject reduction of rewriting in the EPTS, $ \Gamma \vdash B : s $, thus $ B $ is not a top sort. Therefore, by the IH we have $ \trans{\Gamma}  \vdashdk \trans{M} : \El{s}~\trans{B}  $. By the IH applied to $ \Gamma \vdash A : s $ we can show $ \trans{\Gamma} \vdashdk \El{s}~\trans{A} : \Type $, and by \thref{preservation_computation} applied to $ A \equiv B $ we get $ \trans{A} \equiv \trans{B} $. Therefore, it suffices to apply \texttt{Conv} to conclude $ \trans{\Gamma} \vdash \trans{M} : \El{s}~\trans{A} $.
  
  \textbf{Case Abs}: The derivation ends with
  {\small\begin{center}
    \AxiomC{$\Gamma \vdash A : s_{1} $}
    \AxiomC{$\Gamma ,x : A \vdash B :s_{2} $}  
    \AxiomC{$\Gamma , x : A \vdash N : B $}
    \LeftLabel{$(s_{1}, s_{2}, s_3) \in \mathcal{R} $}    
    \RightLabel{\textsc{Abs}}
    \TrinaryInfC{$\Gamma \vdash \lambda_{s_1,s_2}(A, \bind{x} B, \bind{x} N) : \Pi_{s_1,s_2}(A, \bind{x} B) $}
    \DisplayProof
  \end{center}}
By the IH and Lemma \ref{equiv_sort}, we have $\trans{\Gamma} \vdashdk \trans{ A } : \U{s_{1}}$, $\trans{\Gamma}, x : \El{s_1}~\trans{A } \vdashdk \trans{ B } : \U{s_{2}}$ and $\trans{\Gamma}, x : \El{s_1}~\trans{A } \vdashdk \trans{ N } : \El{s_2}~\trans{B}$. By \texttt{Abs} we get $ \trans{\Gamma} \vdashdk \lambda x : \El{s_1}~\trans{A}. \trans{B} :\El{s_1}~\trans{A} \to \U{s_2} $ and $\trans{\Gamma} \vdashdk \lambda x : \El{s_1}~\trans{A }.\trans{ N } : \Pi x : \El{s_{1}}~\trans{A }.\El{s_{2}}~\trans{B}$.

Using inversion of typing, it is not difficult to show that \[
 \trans{\Gamma} \vdashdk  \Pi x : \El{s_{1}}~{A}.\El{s_{2}}~((\lambda x : \El{s_1}~\trans{A} . \trans{ {B} })~x) : \Type
  \,.\]Hence, because $  \Pi x : \El{s_{1}}~{A}.\El{s_{2}}~((\lambda x : \El{s_1}~\trans{A} . \trans{ {B} })~x) \equiv  \Pi x : \El{s_{1}}~\trans{A }.\El{s_{2}}~\trans{B}$, by \texttt{Conv} we can get \[
\trans{\Gamma} \vdashdk \lambda x : \El{s_1}~\trans{A }.\trans{ N } : \Pi x : \El{s_{1}}~{A}.\El{s_{2}}~((\lambda x : \El{s_1}~\trans{A} . \trans{ {B} })~x)
  \,.\]Therefore, we can apply \texttt{Conv} to conclude
\begin{align*}
\trans{\Gamma} &\vdashdk \abs{s_{{1}}}{s_{2}}~\trans{ {A} }~(\lambda x : \El{s_1}~\trans{{A} } . \trans{ {B} }_{\Gamma, x : {A}})~(\lambda x : \El{s_1}~\trans{A }.\trans{ N }) \\ &: \El{s_3}~(\Prod{s_{{1}}}{s_{2}}~\trans{ {A} }~(\lambda x : \El{s_1}~\trans{{A} } . \trans{ {B} }))\qedhere
\end{align*}
}\end{proof}

\section{Conservativity and Adequacy}
\label{sec:conservativity}

Many works proposing \Dedukti{} encodings often stop after showing soundness and leave conservativity as a conjecture. This is because, when mixing the rules $ \beta $ with $ \ctb{beta} $, as done in traditional \Dedukti{} encodings, one needs to show the termination of both, given that to show conservativity one often considers terms in normal form \cite{dowek2007} (with the notable exception of \cite{assaf2015conservativity}). However this problem is non-trivial, and in particular the normalization of $ \beta \cup \ctb{beta} $  implies the termination (and thus normally also the consistency) of the encoded system. This is also unnatural, as logical frameworks should be agnostic to the fact that a system is consistent or not, and thus this shouldn't be required to show conservativity.

In this section we will show how conservativity can be proven without difficulties when we distinguish the rules $ \beta $ and $ \ctb{beta} $. In particular, our proof does not need $ \beta \cup \ctb{beta} $ to be normalizing, and thus also applies to non-normalizing and inconsistent systems.

We start by defining a notion of invertible forms and an inverse translation which allows to invert them into the original system. After proving some basic properties about them, we then proceed with the proof of conservativity.

\subsection{The inverse translation}
\label{subsec:structure}

\begin{definition}[Invertible forms]
  We call the terms generated by the following grammar the \textit{invertible forms}. The $ s_i $ are arbitrary sorts in $ \mathcal{S}$, whereas the $ T_1, T_2 $ are arbitrary terms.
  {\small\begin{align*}
    M, N, A, B ::= &~x \mid \uu{s} \mid \ctb{abs_{s_1,s_2}}~A~(\lambda x : T_1. B)~(\lambda x : T_2. M)\mid (\lambda x : T. M)~N    \\
                   &|~\ctb{Prod_{s_1,s_2}}~A~(\lambda x : T_1. B)\mid \ctb{app_{s_1,s_2}}~A~(\lambda x : T_1. B)~M~N
  \end{align*}}%
\end{definition}

Note that this definition includes some terms which are not in $ \beta $ normal form. The next definition justifies the name of invertible forms: we know how to invert them.

\begin{definition}
  We define the inverse translation function $ \invt{-} : \Lambda_{\textup{\texttt{DK}}} \to \Lambda_{EPTS} $ on invertible forms by structural induction.

{\small\parbox{.2\textwidth}{ 
  \begin{align*}
  &\invt{x} = x\\
    &\invt{\uu{s}} = s\\
  &\invt{(\lambda x : \_. M)~N} = \invt{M} \{ \invt{N}/x \}    
  \end{align*}}
\parbox{.4\textwidth}{ 
  \begin{align*}
  &\invt{\ctb{Prod_{s_1,s_2}}~A~(\lambda x : \_. B)} = \Pi_{s_1,s_2}(\invt{A} , \bind{x} \invt {B})\\
  &\invt{\ctb{abs_{s_1,s_2}}~A~(\lambda x : \_. B)~(\lambda x : \_. M)} = \lambda_{s_1,s_2}(\invt{A}, \bind{x}\invt{B}, \bind{x}\invt{M})\\
  &\invt{\ctb{app_{s_1,s_2}}~A~(\lambda x : \_. B)~M~N} = @_{s_1,s_2}(\invt{A}, \bind{x}\invt{B}, \invt{M}, \invt{N})
\end{align*}}}%
 \end{definition}

We can show, as expected, that the terms in the image of the translation $ \trans{-} $ are invertible forms and that $ \invt{-} $ is a left  inverse of $ \trans{-} $. The proof is a simple induction on $ M $.
 \begin{proposition}
   \label{invt_left_inverse}
   For all $ M \in \Lambda_{EPTS} $, $ \trans{M}$ is an invertible form and  $ \invt{\trans{ M }} = M $.
\end{proposition}

The following lemma shows that invertible forms are closed under rewriting and that this rewriting can also be inverted into the EPTS.

\begin{proposition}
  \label{red_invt}
  Let $ M $ be an invertible form.
  \begin{enumerate}
  \item If $ N $ is an invertible form, then $ M\{N/x\} $ is also and  $ \invt{M} \{ \invt{N}/x \}=\invt{M \{ N/x \}} $.
  \item If $ M \red_{\ctb{beta_{s_1,s_2}}} N $ then $ N $ is an invertible form and $ \invt{M} \red^*_\beta \invt{N}$
  \item If $ M \red_{\beta, \uu{s_1}\text{-red}} N$ then $ N $ is an invertible form and $ \invt{M} = \invt{N} $.
  \item If $ M \red^* N $ then $ N $ is an invertible form and $ \invt{M} \red^* \invt{N} $.
  \end{enumerate}
\end{proposition}
\toshort{\begin{proof}
The  property $ 1 $ is shown by induction on $ M $, whereas $ 2,3 $ follow by induction on the rewrite context and $ 4 $ follows directly from $ 2,3 $.
\end{proof}}
\tolong{\begin{proof}
  \begin{enumerate}
  \item By induction on $ M $.
  \item By induction on the rewrite context. For the base case, we have \[
\app{s_1}{s_2}~A_1~(\lambda x : T_1. B_1)~(\abs{s_1}{s_2}~A_2~(\lambda x : T_2. B_2)~(\lambda x : T_3. M'))~N' \red  (\lambda x : T_3. M')~N'
\]whose right hand side is in the grammar. Moreover, we have \[
@_{s_1,s_2}(\invt{A_1},\bind{x}\invt{B_1}, \lambda_{s_1,s_2}(\invt{A_2},\bind{x} \invt{B_2},\bind{x} \invt{M'}),\invt{N'}) \red \invt{M'} \{ \invt{N'}/x \} = \invt{(\lambda x : T_3. M')~N'} 
\]and thus the reduction is reflected by the inverse translation.
\item By induction on the rewrite context. Note that there is no base case for $ \uu{s_1} $-red, as there is no term of the form $ \El{s_2}~\uu{s_1} $ in the grammar. For the base case of $ \beta $, we have $ (\lambda x : T. M')~N' \red M'  \{N'/x \} $. Hence the resulting term is in the grammar and we have $\invt{(\lambda x : T. M')~N'} = \invt{M'} \{ \invt{N'}/x \} = \invt{M'  \{ N'/x \}}$ by part 1.
\item Immediate consequence of the previous parts.\qedhere
  \end{enumerate}
\end{proof}}

\begin{remark}
Note that this last proposition explains the difference between the $ \beta $ and $ \ctb{beta_{s_1,s_2}} $ steps. Whereas $ \ctb{beta_{s_1,s_2}} $ steps represent the real computation steps that take place in the encoded system, $ \beta $ steps are invisible because they correspond to the framework's substitution, an administrative operation that is implicit in the encoded system. Therefore, it was expected that $ \ctb{beta_{s_1,s_2}} $ steps would be reflected into the original system, whereas $ \beta $ steps would be silent.
\end{remark}

Putting all this together, we deduce that computation and conversion in \Dedukti{} are reflected in the encoded system.

\begin{corollary}[Reflection of computation]\label{conservativity_conversion}
  For $ M, N \in \Lambda_{EPTS} $, we have
  \begin{enumerate}
  \item If  $ \trans{M} \red^* \trans{N} $ then $ M \red^* N $.
  \item If $ \trans{M} \equiv \trans{N} $ then $ M \equiv N $.    
  \end{enumerate}
\end{corollary}
\begin{proof}
  \begin{enumerate}
  \item Immediate consequence of Proposition \ref{red_invt} and Proposition \ref{invt_left_inverse}.
  \item Follows from confluence of  $ \beta\Repts $ and also Proposition \ref{red_invt} and Proposition \ref{invt_left_inverse}.\qedhere
  \end{enumerate}
\end{proof}

Note that for part 2 we really need $ \beta\mathscr{R}_{EPTS} $ to be confluent. Indeed, If $ \trans{M} \invred N $ then we cannot apply $ |-| $ to $ N $ because it might not be an invertible form.
\subsection{Conservativity}
\label{subsec:conservativity}

Before showing conservativity, we show the following auxiliary result, saying that every $ \beta $ normal term $ M $ that has type $ \Pi x : A. B $ in  $ \trans{\Gamma} $ is an abstraction.

\begin{lemma}\label{beta_normal_is_eta_long}
  Let $ M $ be in $ \beta $-normal form. If $ \trans{\Gamma} \vdashdk M : \Pi x : A. B $ then $M = \lambda x : A'. N $ with $ A' \equiv A $ and $\trans{\Gamma}, x : A \vdashdk N : B $. 
\end{lemma}
\begin{proof}
  By induction on $ M $. $ M $ cannot be a variable or constant, as  there is no $ x : C\in \trans{\Gamma} $ or $ c[\Delta] : C \in\Sigmaepts $ with $ C \equiv  \Pi x : A. B $. If $ M = M_1 M_2 $, then $ M_1 $ has a type of the form $ \Pi x' : A'. B' $. By IH we get that $ M_1 $ is an abstraction, which contradicts the fact that $ M $ is in $ \beta $ normal form.

  Therefore, $ M $ is an abstraction, of the form $ M = \lambda x : A'. N $.  By inversion of typing, we thus have $\trans{\Gamma}, x : A' \vdashdk N :B'$ with $ A' \equiv A $ and $ B' \equiv B $. We can then use \thref{dk_conv_in_context} and \texttt{Conv} to derive $ \trans{\Gamma}, x : A \vdashdk N : B  $.
  \end{proof}

  We are now ready to show conservativity for $ \beta $ normal forms. However, if we also want to show adequacy later, we also need to show that $ \invt{-} $ is a kind of right inverse to $ \trans{-} $. But because the inverse translation does not capture the information in the type annotations of binders, $ \trans{\invt{M}} = M $ does not hold.

  \begin{example}
    Take any invertible forms $ A, B $ and a term $ T $ with $ T\neq \El{s_1}~A $. Then the term $ M= \Prod{s_1}{s_2}~A~(\lambda x : T. B) $ is sent by $ \invt{-} $ into $ \Pi_{s_1,s_2}(\invt{A},\bind{x}\invt{B}) $, which is then sent by $ \trans{-} $ into $ \Prod{s_1}{s_2}~\trans{\invt{A}}~(\lambda x : \El{s_1}~\trans{\invt{A}}. \trans{\invt{B}}) $. Therefore, even if have $ \trans{\invt{B}} = B $ and $ \trans{\invt{A}}=A $, we still have $ T \neq \El{s_1}~A $, implying  $ M \neq \trans{\invt{M}} $. However, if $ M $ is typable, then by typing constraints we should nevertheless have $ T \equiv \El{s_1}~A $.
  \end{example}

  Therefore, while proving conservativity we will show a weaker property: for the well-typed terms we are interested in, $ \invt{-} $ is a right inverse up to the following ``hidden'' conversion.

\begin{definition}[Hidden step]
  We say that a rewriting step $ M \red N $ is hidden when it happens on the type annotation of a binder. More formally, we should have a rewriting context $ C(-) $ and terms $ A, A', P $ such that $ A \red A' $, $ M = C(\lambda x : A. P)$ and $ N=C(\lambda x : A'. P) $. We denote the conversion generated by such rules by $ \equiv_H $.
\end{definition}

We now have all ingredients to show that the encoding is conservative for $ \beta $ normal forms.

\begin{theorem}[Conservativity of $ \beta $ normal forms]
  \label{conservativity_beta_normal_forms}
  Suppose  $ \Gamma \vdash A~type $ and let $ M \in \Lambdadk$ be a $ \beta $ normal form such that $\trans{\Gamma} \vdashdk M : T$, with $T = \El{s_A}~\trans{A }$ or $ T=\U{A} $. Then  $ M$ is an invertible form, $ \Gamma \vdash \invt{M} : A $ and $ \trans{ \invt{M} } \equiv_H M $.
\end{theorem}
\begin{proof}
By induction on $ M $.

\textbf{Case $ M = \lambda x : A'. M' $} : By inversion we have $ M : \Pi x : A'_1. A'_2 $ with $ T \equiv \Pi x : A'_1. A'_2 $. This then implies that $ T $ reduces to a dependent product, but because $ T $ is of the form $\El{s_A}~\trans{A }$ or $ \U{A} $ and $ \mathscr{R}_{\texttt{EPTS}} $ is arity preserving, this cannot hold. Thus, this case is impossible.

  \textbf{Case $ M = M_1M_2 $} : As $ M $ is in beta normal form, its head symbol is a constant or variable. However, there is no $ c[\Delta] :C \in \Sigma_{\texttt{EPTS}} $ or $ x : C\in \Gamma$ with $ C $ convertible to a dependent product type. Hence, this case is impossible.
  
  \textbf{Case $ M = x $} : If $ M = x $, by inversion of typing there is $ x : \El{s_B}~\trans{B} \in \trans{\Gamma} $ with $ T \equiv \El{s_B}~\trans{B}$. Therefore, we deduce $ A \equiv B $ and thus we can  derive $ \Gamma \vdash x : A   $ by applying \textsc{Var} with $ x : B \in \Gamma $, then \textsc{Conv} with $ A \equiv B $ and $ \Gamma \vdash A~type $.

  \textbf{Case $ M = c[\vec{M}]$} : We proceed by case analysis on $ c $. \tolong{Note that for $ c = \El{s}~M' $ or $ c = \U{s} $ the resulting type is $ \Type $, which is not convertible to $ T $. Hence, these cases are impossible.}\toshort{We present only case $ c = \Prod{s_1}{s_2} $ here and refer to the long version in \cite{adequate} for all the details.}
  
  \begin{note}
    In the following, to improve readability we omit the typing hypothesis when applying \texttt{Conv}. However, all such uses can be justified.
  \end{note}

\tolong{  \textbf{Case $ c = \uu{s_1} $}: As we have $ \trans{\Gamma} \vdashdk \uu{s_1} : \U{s_2} $, by uniqueness of types we have $ T \equiv \U{s_2}$, and therefore we get $ A \equiv s_2 $. We can thus deduce $ \Gamma \vdash s_1 : A $ by using \textsc{Conv} with $ \Gamma \vdash s_1 : s_2 $.}
  
  \textbf{Case $ c= \Prod{s_1}{s_2}$}: By inversion of typing, we have

\begin{enumerate}
  \item $ \vec{M} = M_1~M_2 $
  \item $\trans{\Gamma} \vdashdk M_1 : \U{s_1} $
  \item $\trans{\Gamma} \vdashdk M_2 : \El{s_1}~M_1 \to \U{s_2} $
  \item $ T \equiv \U{s_3} $
  \end{enumerate}

As $ M_1 $ is in $ \beta $ normal form, by IH $ M_1$ is an invertible form, $ \Gamma \vdash \invt{M_1} : s_1 $ and $ \trans{ \invt{M_1} } \equiv_{H} M_1 $.
  
By Lemma \ref{beta_normal_is_eta_long} applied to 3, we get $ M_2 = \lambda x : B. N $ and $ B \equiv \El{s_1}~M_2 $ with $ \trans{\Gamma}, x:\El{s_1}~M_1 \vdash N :  \U{s_2} $. Because $ M_1 \equiv \trans{\invt{M_1}} $, we have $ \trans{\Gamma}, x:\El{s_1}~\trans{\invt{M_1}} \vdash N :  \U{s_2} $. As $ \Gamma \vdash \invt{M_1} : s_1 $ we have $ \Gamma, x : \invt{M_1}~\WF $ and thus by IH $ N $ is an invertible form and we have $ \trans{\invt{N}} \equiv_H N $ and $ \Gamma, x:\invt{M_1} \vdash \invt{N} :s_2 $.

Therefore, by \textsc{Prod} we have $ \Gamma \vdash \Pi_{s_1,s_2}(\invt{M_1},\bind{x} \invt{N}) : s_3 $, and then by \textsc{Conv} with $ A \equiv s_3 $ we conclude $ \Gamma \vdash \Pi_{s_1,s_2}(\invt{M_1},\bind{x} \invt{N}) : A $. Finally, as $ M_1 \equiv_H \trans{\invt{M_1}} $, $ N \equiv_H\trans{\invt{N}} $ and $ B \equiv \El{s_1}~M_1 \equiv \El{s_1}~\trans{\invt{M_1}} $, we conclude
\begin{align*}
M &=\Prod{s_1}{s_2}~M_1~M_2 =\Prod{s_1}{s_2}~M_1~(\lambda x : B.N) \\ &\equiv_H \Prod{s_1}{s_2}\trans{\invt{M_1}}~(\lambda x : \El{s_1}~\trans{\invt{M_1}}.\trans{\invt{N}}) = \trans{\Pi_{s_1,s_2}(\invt{M_1},\bind{x} \invt{N})}= \trans{\invt{M}}\toshort{\qedhere}
\end{align*}

     \tolong{\textbf{Case $ c = \abs{s_1}{s_2} $}: By inversion of typing, we have

       \begin{enumerate}
  \item $ \vec{M} = M_1~M_2~M_3$
\item $ \trans{\Gamma} \vdash M_1 : \U{s_1} $
\item $ \trans{\Gamma} \vdash M_2 : \El{s_1}~M_2 \to \U{s_2} $
\item $ \trans{\Gamma} \vdash M_3 : \Pi x : \El{s_1}~M_1. \El{s_2}~(M_2~x) $
\item $ T \equiv \El{s_3}~(\Prod{s_1}{s_2}~M_1~M_2) $
\end{enumerate}

By the same arguments as in case $ M = \Prod{s_1}{s_2}~\vec{M} $, we have that

\begin{itemize}
\item $ M_1 $ is an invertible form, $ \trans{ \invt{M_1} } \equiv_{H} M_1 $ and $ \Gamma \vdash \invt{M_1} : s_1 $.
\item $ M_2 = \lambda x : B. N $, $ N $ is an invertible form, $ \trans{\invt{N}} \equiv_H N $, $ \lambda x : \El{s_1}~\trans{\invt{M_1}}. \trans{ \invt{N} } \equiv_{H} \lambda x : B. N $ and $ \Gamma, x : \invt{M_1} \vdash \invt{N} : s_2 $.
\end{itemize}

By Lemma \ref{beta_normal_is_eta_long} applied to 4 we have $ M_3 = \lambda x : C. P $, $ C \equiv \El{s_1}~M_1 $ and $ \trans{\Gamma}, x : \El{s_1}~M_1 \vdashdk P : \El{s_2}~(M_2~x) $. Using $ M_2 = \lambda x : B. N $ and \thref{reduce_type}, we get $\trans{\Gamma}, x : \El{s_1}~M_1 \vdashdk P : \El{s_2}~N$. Because $ M_1 \equiv \trans{\invt{M_1}} $ and $ N \equiv \trans{\invt{N}} $, we then get $ \trans{\Gamma}, x : \El{s_1}~\trans{\invt{M_1}} \vdashdk P : \El{s_2}~\trans{\invt{N}} $.

Therefore, by IH  $ P $ is an invertible form, $ \Gamma, x : \invt{M_1} \vdash \invt{P} : \invt{N} $ and $ \trans{\invt{P}} \equiv_H P $. Putting this together with $ \Gamma \vdash \invt{M_1} : s_1 $ and $ \Gamma, x : \invt{M_1} \vdash \invt{N} : s_2 $ we can derive $ \Gamma \vdash \lambda_{s_1,s_2}(\invt{M_1}, \bind{x}\invt{N}, \bind{x}\invt{P}) : \Pi_{s_1,s_2}(\invt{M_1}, \bind{x}\invt{N}) $. From 5 we can also show $ \Pi_{s_1,s_2}(\invt{M_1}, \bind{x}\invt{N}) \equiv A $, which allows us to apply \textsc{Conv} to get $ \Gamma \vdash \lambda_{s_1,s_2}(M_1, \bind{x}\invt{N}, \bind{x}\invt{P}) : A $. Finally, from $ \trans{ \invt{M_1} } \equiv_{H} M_1 $, $ \lambda x : B. N \equiv_H \lambda x : \El{s_1}~\trans{\invt{M_1}}. \trans{ \invt{N} }$,  $ P \equiv_H \trans{ \invt{P} } $ and $ C \equiv \El{s_1}~M_1 \equiv \El{s_1}~\trans{\invt{M_1}} $ we get
\begin{align*}
&M = \abs{s_1,s_2}~M_1~M_2~M_3  = \abs{s_1,s_2}~M_1~(\lambda x:B.N)~(\lambda x : C.P)\\ &\equiv_H \abs{s_1,s_2}~\trans{\invt{M_1}}~(\lambda x : \El{s_1}~\trans{\invt{M_1}}. \trans{\invt{N}})~(\lambda x : \El{s_1}~\trans{\invt{M_1}}. \trans{\invt{P}})\\ &= \trans{\lambda_{s_1,s_2}(\invt{M_1}, \bind{x}\invt{N}, \bind{x}\invt{P})} = \trans{\invt{M}}
\end{align*}

\textbf{Case $ c =  \app{s_1}{s_2} $}: By inversion of typing, we have

\begin{enumerate}
\item $ \vec{M} = M_1~M_2~M_3~M_4~M_5 $
\item $ \trans{\Gamma} \vdashdk M_1 : \U{s_1} $
\item $ \trans{\Gamma} \vdashdk M_2 : \El{s_1}~M_1 \to \U{s_2} $
\item $ \trans{\Gamma} \vdashdk M_3 : \El{s_3}~(\Prod{s_1}{s_2}~M_1~M_2) $
\item $ \trans{\Gamma} \vdashdk M_4 : \El{s_1}~M_1$
\item $ T \equiv \El{s_2}~(M_2~M_4) $
\end{enumerate}

By the same arguments as in case $ M = \Prod{s_1}{s_2}~\vec{M} $, we have that

\begin{itemize}
\item $ M_1 $ is an invertible form, $ \trans{ \invt{M_1} } \equiv_{H} M_1 $ and $ \Gamma \vdash \invt{M_1} : s_1 $.
\item $ M_2 = \lambda x : B. N $, $ N $ is an invertible form, $ \trans{\invt{N}} \equiv_H N $, $ \lambda x : \El{s_1}~\trans{\invt{M_1}}. \trans{ \invt{N} } \equiv_{H} \lambda x : B. N $ and $ \Gamma, x : \invt{M_1} \vdash \invt{N} : s_2 $.
\end{itemize}

As $ \Prod{s_1}{s_2}~M_1~M_2 \equiv \trans{\Pi_{s_1,s_2}(\invt{M_1},\bind{x}\invt{M_2})} $, from 4 we get $ \trans{\Gamma} \vdashdk M_3 : \El{s_3}~\trans{\Pi_{s_1,s_2}(\invt{M_1},\bind{x}\invt{M_2})} $. Therefore, we deduce by the IH that $ M_3 $ is an invertible form, $ \Gamma \vdash \invt{M_3} : \Pi_{s_1,s_2}(\invt{M_1}, \bind{x}\invt{N}) $ and $ \trans{\invt{M_3}} \equiv_H M_3 $.

Moreover, as $ M_1 \equiv \trans{\invt{M_1}} $, from 5 we  get $ \trans{\Gamma}\vdashdk M_4 :\El{s_1}~\trans{\invt{M_1}} $, therefore by IH we deduce that $ M_4 $ is an invertible form, $ \Gamma \vdash \invt{M_4} : \invt{M_1} $ and $ \trans{\invt{M_4}} \equiv_H M_4 $.

Putting together $ \Gamma \vdash \invt{M_1} : s_1 $, $ \Gamma, x : \invt{M_1} \vdash \invt{N} : s_2 $, $ \Gamma \vdash \invt{M_3} : \Pi_{s_1,s_2}(\invt{M_1}, \bind{x}\invt{N}) $ and $ \Gamma \vdash \invt{M_4} : \invt{M_1} $ we  derive $ \Gamma \vdash @_{s_1,s_2}(\invt{M_1},\bind{x}\invt{N},\invt{M_3},\invt{M_4}) : \invt{N}(\invt{M_4}/x) $.

From 8 we get $ T \equiv \El{s_2}~(M_2~M_4) \equiv \El{s_2}~((\lambda x : \El{s_1}~\trans{\invt{M_1}}. \trans{\invt{N}})~\trans{\invt{M_4}}) \equiv \El{s_2}~\trans{\invt{N}  \{ \invt{M_4}/x \}} $, thus we deduce $ A \equiv \invt{N} \{\invt{M_4}/x \} $. Hence, we can apply \textsc{Conv} to get $ \Gamma \vdash @_{s_1,s_2}(\invt{M_1},\bind{x}\invt{N},\invt{M_3},\invt{M_4}) : A $.

From $ \trans{ \invt{M_1} } \equiv_{H} M_1 $, $ \lambda x : \El{s_1}~\trans{\invt{M_1}}. \trans{ \invt{N} } \equiv_{H} \lambda x : B. N $, $  \trans{\invt{M_4}} \equiv_H M_4  $ and $ \trans{\invt{M_3}} \equiv_H M_3 $ we can then conclude
\begin{align*}
&M =  \app{s_1}{s_2}~M_1~M_2~M_3~M_4 = \app{s_1}{s_2}~M_1~(\lambda x:B.N)~M_3~M_4 \\ &\equiv_H  \app{s_1}{s_2}~\trans{\invt{M_1}}~(\lambda x : \El{s_1}~\trans{\invt{M_1}}. \trans{ \invt{N} })~\trans{\invt{M_3}}~\trans{\invt{M_4}}\\ &=\trans{@_{s_1,s_2}(\invt{M_1},\bind{x}\invt{N},\invt{M_3},\invt{M_4})} = \trans{\invt{M}}\qedhere
\end{align*}}\end{proof}

By \thref{basic-properties}, $\beta $ is strongly normalizing and type preserving. Therefore from the previous result we can immediately get full conservativity.

\begin{theorem}[Conservativity]
  \label{conservativity}
Let $ \Gamma \vdash A~type $, $ M \in \Lambdadk$ such that $ \trans{\Gamma} \vdashdk M : T $, with $T = \El{s_A}~\trans{A }$ or $ T= \U{A} $. We have $ \Gamma \vdash \invt{NF_{\beta}(M)} : A  $ and $ M \red^*_{\beta} NF_{\beta}(M)\equiv_{H}\trans{\invt{NF_{\beta}(M)}} $.
\end{theorem}

Note that this also gives us a straightforward algorithm to invert terms: it suffices to  normalize with $ \beta $ and then  apply $ \invt{-} $.

\subsection{Adequacy}
\label{subsec:adeq}

If we write $ \Lambda(\Gamma \vdash_{EPTS} \_ : A) $ for the set of $ M \in \Lambda_{EPTS} $ such that $\Gamma \vdash M : A $  and $ \Lambda_{NF}(\Gamma \vdashdk \_ : T) $ for the set of $ M \in \Lambdadk $ in $ \beta$ normal form such that $\Gamma \vdashdk M : T $, we can show our adequacy theorem. This result follows by simply putting together  \thref{basic-properties}, \thref{preservation_computation}, \thref{soundness}, \thref{conservativity_conversion} and \thref{conservativity}.

\begin{theorem}[Computational adequacy]
  For $ A, \Gamma $ with $ \Gamma \vdash A~type $, let $ T = \U{A} $ if $ A $ is a top sort, otherwise $ T = \El{s_A}~\trans{A} $. We have a bijection \[
     \Lambda(\Gamma \vdash_{EPTS} \_ : A) \simeq \Lambda_{NF}(\trans{\Gamma} \vdashdk \_ : T) /\equiv_{H}
 \]given by $ \trans{-} $ and $ \invt{-} $. It is compositional in the sense that $ \trans{-}$ commutes with substitution. It is computational in the sense that $ M \red^* N $ iff $ \trans{M} \red^* \trans{N} $. Moreover, any $ M $ satisfying $ \trans{\Gamma} \vdashdk M : T $ has such a $ \beta $ normal form.
\end{theorem}

\section{Representing systems with infinitely many sorts}
\label{sec:representation_sorts}

We have presented an encoding of EPTSs in \Dedukti{} that is sound, conservative and adequate. However when using it in practice with \Dedukti{} implementations we run into problems when representing systems with infinitely many sorts, such as in Martin-L\"of's Type Theory or the Extended Calculus of Constructions. Indeed, in this case our encoding needs an infinite number of constant and rule declarations, which cannot be made in practice.

One possible solution is to approximate the infinite sort structure by a finite one. Indeed, every proof in an infinite sort systems only uses a finite number of sorts, and thus does not need all of them to be properly represented.

A different approach proposed in \cite{assaf} is to internalize the indices of $ \Prod{s_1}{s_2}, \El{s_1}, ... $ and represent them inside \Dedukti{}. In order to apply this method, we  chose to stick with systems in which $ \mathcal{A}, \mathcal{R} $ are total functions $ \mathcal{S} \to \mathcal{S} $ and $ \mathcal{S} \times \mathcal{S} \to \mathcal{S} $ respectively. Note that this is true for almost all infinite sort systems used in practice, and this will greatly simplify our presentation.

We can now declare a constant $ \ctb{\hat{\mathcal{S}}} $  to represent the type of sorts in $ \mathcal{S} $ and two constants $ \ctb{\hat{\mathcal{A}}}, \ctb{\hat{\mathcal{R}}} $ to represent the functions $ \mathcal{A}, \mathcal{R} $. Then, each of  our previously declared families of constants now becomes a single one, by  taking arguments of type $ \ctb{\hat{\mathcal{S}}} $. The same happens with the rewrite rules. This leads to the theory presented in Figure \ref{theory-sort}, which we call $ (\Sigmaepts^{S}, \Repts^S ) $.

\begin{figure}[h]
{\small\noindent\parbox{.5\textwidth}{
  \begin{align*}
    &\ctb{\hat{\mathcal{S}}} : \Type\\
    &\ctb{\hat{\mathcal{A}}} [s_1 : \ctb{\hat{\mathcal{S}}}] : \ctb{\hat{\mathcal{S}}}\\
    &\ctb{\hat{\mathcal{R}}} [s_1 : \ctb{\hat{\mathcal{S}}}; s_2 :  \ctb{\hat{\mathcal{S}}}] : \ctb{\hat{\mathcal{S}}}
\end{align*}}
\parbox{.5\textwidth}{
\begin{align*}
  &\ctb{U} [s : \ctb{\hat{\mathcal{S}}}] : \Type \\
  &\ctb{El}[s: \ctb{\hat{\mathcal{S}}}; A: \ctb{U}~s] : \Type \\    
    &\uu{} [s : \ctb{\hat{\mathcal{S}}}] : \ctb{U}~(\ctb{\hat{\mathcal{A}}}~s) \\
    &\ctb{El}~s'~(\uu{}~s) \red_{\uu{}\text{-red}}    \U{}~s 
\end{align*}}
\noindent\begin{align*}
&\ctb{Prod} [s_1 : \ctb{\hat{\mathcal{S}}}; s_2 :  \ctb{\hat{\mathcal{S}}}; A : \U{}~s_1; B : \El{}~s_1~A \to \U{}~s_2] : \U{}~(\ctb{\hat{\mathcal{R}}}~s_1~s_2) &\\
  &\ctb{abs} [s_1 : \ctb{\hat{\mathcal{S}}}; s_2 :  \ctb{\hat{\mathcal{S}}}; A : \U{}~s_1; B : \El{}~s_1~A \to \U{}~s_2; N : \Pi x: \El{}~s_1~A. \El{}~s_2~(B~x)] \\
  &\hspace{26em} :  \El{}~(\ctb{\hat{\mathcal{R}}}~s_1~s_2)~(\ctb{Prod}~s_1~s_2~A~B)\\
  &\ctb{app} [s_1 : \ctb{\hat{\mathcal{S}}}; s_2 :  \ctb{\hat{\mathcal{S}}}; A : \U{}~s_1; B : \El{}~s_1~A \to \U{}~s_2; M : \El{}~(\ctb{\hat{\mathcal{R}}}~s_1~s_2)~(\ctb{Prod}~s_1~s_2~A~B); N : \El{}~s_1~A] \\
  &\hspace{26em} :  \El{}~s_2~(B~N)\\
  &\ctb{app}~s_1~s_2~A~B~(\ctb{abs}~s_1'~s_2'~A'~B'~M)~N \red_{\ctb{beta}} M~N
         \end{align*}}
       \\[-2\baselineskip]
       \caption{Definition of the theory $ (\Sigmaepts^{S}, \Repts^S )$}
       \label{theory-sort}
\end{figure}

\newcommand{\senc}[1]{\dot{#1}}

This theory needs of course to be completed case by case, so that $ \ctb{\hat{\mathcal{S}}},\ctb{\hat{\mathcal{A}}} , \ctb{\hat{\mathcal{R}}} $ correctly represent $ \mathcal{S},\mathcal{A}, \mathcal{R} $. For this to hold, each sort  $ s \in \mathcal{S} $ should have a representation $ \senc{s} : \ctb{\hat{\mathcal{S}}}$, and this should restrict to a bijection when considering only the closed normal forms of type $ \ctb{\hat{\mathcal{S}}} $. Moreover, we should add rewrite rules such that $ \mathcal{A}(s_1) = s_2 $ iff $ \ctb{\hat{\mathcal{A}}}~\senc{s_1} \equiv \senc{s_2} $ and $ \mathcal{R}(s_1,s_2) = s_3 $ iff $ \ctb{\hat{\mathcal{R}}}~\senc{s_1}~\senc{s_2} \equiv \senc{s_3} $.

In order to understand intuitively these conditions, let's look at an example.

\begin{example}
  The sort structure of Martin-L\"of's Type Theory is given by the specification $ \mathcal{S} = \mathbb{N} $, $ \mathcal{A}(x) = x + 1 $ and $ \mathcal{R}(x,y) = max\{x,y\} $. We can represent this in \Dedukti{} by declaring constants $ \ctb{z} :  \ctb{\hat{\mathcal{S}}}$, $ \ctb{s} [n : \ctb{\hat{\mathcal{S}}}] : \ctb{\hat{\mathcal{S}}}$ and rewrite rules $ \ctb{\hat{\mathcal{A}}}~x \red \ctb{s}~x, \ctb{\hat{\mathcal{R}}}~\ctb{z}~x \red x,\ctb{\hat{\mathcal{R}}}~x~\ctb{z} \red x$ and $ \ctb{\hat{\mathcal{R}}}~(\ctb{s}~x)~(\ctb{s}~y) \red \ctb{s}~(\ctb{\hat{\mathcal{R}}}~x~y) $. 
\end{example}

\tolong{
With this representation, we can revisit the example of the polymorphic identity function.

\begin{example}
The (predicative and at sort $ 0 $) polymorphic identity function in Martin L\"of's Type Theory is given by the term \[
\lambda_{1,0}(0, \bind{A} \Pi_{0,0}(A,\bind{x}A), \bind{A} \lambda_{0,0}(A, \bind{x}A,\bind{x}x))
  \,.\]It can be represented in the encoding by \[
\ctb{abs}~(\ctb{s~z})~\ctb{z}~(\ctb{u}~\ctb{z})~(\lambda A. \ctb{Prod}~\ctb{z}~\ctb{z}~A~(\lambda x.A))~(\lambda A. \ctb{abs}~\ctb{z}~\ctb{z}~A~(\lambda x. A)~(\lambda x.x))
  \,.\]
\end{example}

Let us now define the encoding function formally, by the following equations.
{\small\begin{align*}
  &\trans{ x }_S = x\\
  &\trans{ s }_S = \ctb{u}~\senc{s} \\
  &\trans{ \Pi_{s_1,s_2}(A, \bind{x} B) }_S = \ctb{Prod}~\senc{s}_1~\senc{s_2}~\trans{ A }_S~(\lambda x : \ctb{El}~\senc{s_1}~\trans{A}_S. \trans{ B }_S)\\
  &\trans{ \lambda_{s_1,s_2}( A, \bind{x} B, \bind{x} M) }_S = \ctb{abs}~\senc{s}_1~\senc{s_2}~\trans{ A }_S~(\lambda x : \ctb{El}~\senc{s_1}~\trans{A}_S. \trans{ B }_S)~(\lambda x : \ctb{El}~\senc{s_1}~\trans{A}_S. \trans{ M }_S)\\
         &\trans{ @_{s_1,s_2}( A, \bind{x} B,  M,  N) }_S = \ctb{app}~\senc{s}_1~\senc{s_2}~\trans{ A }_S~(\lambda x : \ctb{El}~\senc{s_1}~\trans{A}_S. \trans{ B }_S)~\trans{ M }_S~\trans{ N }_S\\
         &\trans{-}_S = -\\
         &\trans{\Gamma, x : A}_S = \trans{\Gamma}_S, x : \ctb{El}~\senc{s_A}~\trans{A}_S\quad\text{where } \Gamma \vdash A : s_A         
\end{align*}}%
}

Now one can proceed as before with the proofs of soundness, conservativity and adequacy, which  follow the same idea as the previously presented ones. However, it is quite unsatisfying that we have to redo all the work of Sections \ref{sec:soundness} and \ref{sec:conservativity} another time, and therefore one can wonder if we can reuse the results we already have about the first encoding.

\toshort{Note that one may intuitively think of the $ (\Sigmaepts^{S}, \Repts^S ) $ as a ``hidden implementation'' of $ (\Sigmaepts, \Repts ) $. In this case, it should be possible to take a proof written in the $ (\Sigmaepts, \Repts ) $ and  ``implement'' it in the $ (\Sigmaepts^{S}, \Repts^S ) $. Following this intuition, we define in the long version in \cite{adequate} a notion of \textit{theory morphism} which allows us to show the soundness of this new encoding by means of morphism from $ (\Sigmaepts, \Repts ) $ to $ (\Sigmaepts^{S}, \Repts^S ) $.

  Nevertheless, this definition has too strong requirements and cannot be used to define a morphism in the other direction to show conservativity. Therefore, it is still an open problem for us to find a notion of morphism  allowing to show  the equivalence between the encodings. For the time being, in order to show conservativity (and then adequacy) of this new encoding one  has to redo the work of Section \ref{sec:conservativity}.}

\tolong{Note that one may intuitively think of the $ (\Sigmaepts^{S}, \Repts^S ) $ as a ``hidden implementation'' of $ (\Sigmaepts, \Repts ) $. In this case, it should be possible to take a proof written in the $ (\Sigmaepts, \Repts ) $ and  ``implement'' it in the $ (\Sigmaepts^{S}, \Repts^S ) $. To formalize this intuition, we will define  a notion of \textit{theory morphism} which will  allows us to establish the soudness of this new encoding using a morphism from $ (\Sigmaepts, \Repts ) $ to $ (\Sigmaepts^{S}, \Repts^S ) $. 

\section{Theory morphisms}
\label{subsec:th}

To define our notion of theory morphism, we start by defining an auxiliary weaker notion of pre-morphism. In the following, we write $ \mathcal{C}(\Sigma_i) $ for the constants appearing in $ \Sigma_i $ and $ \Lambda(\Sigma_i) $ for the terms built using such constants.

  \begin{definition}[Theory pre-morphism]
    A \textit{theory pre-morphism} $ F : (\Sigma_1,\mathscr{R}_1) \to (\Sigma_2, \mathscr{R}_2) $ is for each $ c \in \mathcal{C}(\Sigma_1) $ a term $ F_c \in \Lambda(\Sigma_2) $ with free variables in $ \Delta_c $.
    Each such $ F $ defines a map on terms $ |-|_F $ given by

    \parbox{.5\textwidth}{
      \begin{align*}
        |c[\vec{M}]|_F &= F_c\{|\vec{M}|_F\} \\
        |x|_F &=x\\
        |\Type|_F &=\Type\\
        |\Kind|_F &= \Kind
      \end{align*}}    
\parbox{.5\textwidth}{
  \begin{align*}
    |\Pi x:A.B|_F &= \Pi x : |A|_F . |B|_F\\
    |\lambda x : A. M|_F &= \lambda x : |A|_F . |M|_F\\
    |M N|_F &= |M|_F |N|_F
  \end{align*}}
\end{definition}

  Given a term $ c[\vec{M}] $ defined in the signature $ \Sigma_1 $, one should understand $  F_c\{|\vec{M}|_F\}  $ as the implementation in $ \Sigma_2 $ of this term. With this interpretation, we can see $ F_c $ as the body of the implementation. This also explains why $ F_c $ should have free variables in $ \Delta_c $, as these corresponds to the arguments  that are supplied to $ c $.

  In order to understand intuitively the definition, let's define a theory pre-morphism from $ (\Sigmaepts, \Repts ) $ to $ (\Sigmaepts^{S}, \Repts^S ) $, which will then be used to show soundness of $ \trans{-}_S $.

  \begin{example}
    We define the pre-morphism $ \phi : (\Sigmaepts, \Repts ) \to (\Sigmaepts^{S}, \Repts^S ) $ by the following data.   We recall in the right the variables in the  context of each constant (we write $ \mathcal{V}(\Delta) $ for the variables in $ \Delta $).
   \begin{align*}
          &\phi_{\U{s}} = \ctb{U}~\senc{s} &&\mathcal{V}(\Delta_{\U{s}}) = -\\
          &\phi_{\El{s}} = \ctb{El}~\senc{s}~A &&\mathcal{V}(\Delta_{\El{s}}) = A\\
          &\phi_{\uu{s}} = \ctb{u}~\senc{s} &&\mathcal{V}(\Delta_{\uu{s}}) = -\\
          &\phi_{\Prod{s_1}{s_2}} = \ctb{Prod}~\senc{s_1}~\senc{s_2}~A~B &&\mathcal{V}(\Delta_{\Prod{s_1}{s_2}}) = A, B\\
          &\phi_{\abs{s_1}{s_2}} = \ctb{abs}~\senc{s_1}~\senc{s_2}~A~B~N &&\mathcal{V}(\Delta_{\abs{s_1}{s_2}}) = A, B, N\\
          &\phi_{\app{s_1}{s_2}} = \ctb{app}~\senc{s_1}~\senc{s_2}~A~B~M~N&&\mathcal{V}(\Delta_{\app{s_1}{s_2}}) = A, B, M, N
      \end{align*}
    We can then calculate for instance the value of $ |\Prod{s_1}{s_2}~T_1~T_2|_\phi $ as \[
      |\Prod{s_1}{s_2}~T_1~T_2~|_\phi = (\ctb{Prod}~\senc{s_1}~\senc{s_2}~A~B) \{|T_1|_\phi/A, |T_2|_\phi/B\} = \ctb{Prod}~\senc{s_1}~\senc{s_2}~|T_1|_\phi~|T_2|_\phi
      \]More generally, we can prove that $ |\trans{M}|_{\phi}= \trans{M}_S $ by induction on $ M \in \Lambda_{EPTS} $.  
  \end{example}

  Not every theory pre-morphism should be called a morphism, as there are some properties which one should enforce. In the following, we write $ \vdash_i $ for a judgment in the theory $ (\Sigma_i,\mathcal{R}_i) $.
  
  \begin{definition}[Theory morphism]\label{morphism}
    A \textit{theory morphism} $ F : (\Sigma_1,\mathscr{R}_1) \to (\Sigma_2, \mathscr{R}_2) $ is a theory pre-morphism satisfying the following conditions
    \begin{enumerate}
    \item for all $ c[A_c] : \Delta_c \in \Sigma_1 $, we have $ |\Delta_c|_F \vdash_2 F_c : |A_c|_F$
    \item for all $ l \red_1 r \in \mathscr{R}_1 $ we have $ |l|_F \red^*_2 |r|_F $
    \end{enumerate}
  \end{definition}

We have the following basic properties about compositionality and preservation of computation and of conversion.

  \begin{lemma}\label{morphism_reduction}
    For each morphism $ F $, we have the following properties.
    \begin{enumerate}
    \item Compositionality: $ |M|_F\{|N|_F/x\} = |M\{N/x\}|_F $
    \item Preservation of computation: if $ M \red_1 N $ then $ |M|_F \red^*_2 |N|_F $
    \item Preservation of conversion: if $ M \equiv_1 N $ then $ |M|_F \equiv_2 |N|_F $
    \end{enumerate}
  \end{lemma}

We can now show the main result about theory morphisms.
  
  \begin{theorem}[Preservation of typing]\label{pres_typing}
    Let $ F : (\Sigma_1,\mathscr{R}_1) \to (\Sigma_2, \mathscr{R}_2)  $ be a theory morphism. 
    \begin{enumerate}
    \item If $ \Gamma~\textup{\texttt{well-formed}}_1$ then $ \vdash_2 |\Gamma|_F~\textup{\texttt{well-formed}}_2 $
    \item If $ \Gamma \vdash_1 M : A $ then $ |\Gamma|_F \vdash_2 |M|_F : |A|_F $
    \end{enumerate}
  \end{theorem}
  \begin{proof}
    By induction on the judgment tree. We do only cases \texttt{Conv} and \texttt{Cons}, as they are the only interesting ones.

    \textbf{Case Cons}: The proof ends with
    
\begin{center}
\AxiomC{$\Delta_c \vdash_1 A_c : s $}
\AxiomC{$\Gamma \vdash_1 \vec{M} : \Delta_c $}  
\RightLabel{\texttt{Cons}}
\LeftLabel{$c [\Delta_c] : A_c \in \Sigma_1$}
\BinaryInfC{$\Gamma \vdash_1 c[\vec{M}] : A_c\{\vec{M}\} $}
\DisplayProof
\end{center}

By IH we have  $ |\Gamma| \vdash_2 |\vec{M}| : |\Delta_c| $. Moreover, because $ F $ is a morphism we have $ |\Delta_c| \vdash_2 F_c : |A_c| $. By substitution we thus deduce $ |\Gamma| \vdash_2 F_c \{|\vec{M}|\} : |A_c|\{|\vec{M}|\} $. Finally, as $ |A_c|\{|\vec{M}|\} = |A_c\{\vec{M}\} |$ we get the result.

\textbf{Case Conv:} The proof ends with

\begin{center}
  \AxiomC{$\Gamma \vdash_1 M : A $}
  \AxiomC{$\Gamma \vdash_1 B : s $}  
  \RightLabel{\texttt{Conv}}
  \LeftLabel{$A \equiv_1 B$}  
\BinaryInfC{$\Gamma \vdash_1 M : B $}
\DisplayProof
\end{center}

By the IH, we have $|\Gamma| \vdash_2 |M| : |A| $ and $|\Gamma| \vdash_2 |B|: s $. Moreover, as $ A \equiv_1 B $, by Lemma \ref{morphism_reduction} we have $ |A| \equiv_2 |B| $, and thus we can apply \texttt{Conv} to conclude.
\end{proof}

\newcommand{\Th}{\textup{\textbf{Th}}}

Using this result, one can also show, as expected, that theories and their morphisms assemble into a category. However, as we will not need this result here, we will not show it. Instead, let's now come back to our pre-morphism $ \phi $ and show that it is indeed a morphism.

    \begin{example}We show that $ \phi $ verifies the conditions of Definition \ref{morphism}, and is thus a morphism. Condition 2 can be easily verified, so we concentrate in the first one.  As an example, we show the propertiy only for constant $ \Prod{s_1}{s_2} $. We need to show \[
        A : |\U{s_1}|_\phi, B : |\El{s_1}~A|_\phi \vdash \ctb{U}~\senc{s_1}~\senc{s_2}~A~B : |\U{s_3}|
      \]where $ (s_1,s_2,s_3) \in \mathcal{R} $. Because $ |\U{s_1}|_\phi = \ctb{U}~\senc{s_1}$ and $  |\El{s_1}~A|_\phi= \ctb{El}~\senc{s_1}~A $ we can show, using rule \texttt{Cons}, that $         A:\ctb{U}~\senc{s_1}, B: \ctb{El}~\senc{s_1}~A \vdash \ctb{U}~\senc{s_1}~\senc{s_2}~A~B : \ctb{U}~(\ctb{\hat{\mathcal{R}}}~\senc{s_1}~\senc{s_2})$. However, as we have $ \ctb{\hat{\mathcal{R}}}~\senc{s_1}~\senc{s_2} \equiv \senc{s_3} $ and $ \ctb{U}~\senc{s_3} : \Type $, using \texttt{Conv} we deduce the required result.
    \end{example}

    We can now use this to show that $ \trans{-}_S $ is sound. Indeed, because $ \trans{-} $ is sound and we have $ |\trans{M}|_\phi = \trans{M}_S $, by Theorem \ref{pres_typing} we immediately get the following result.
    
    \begin{corollary}[$ \trans{-}_S $ is sound]\label{soudness2}
  Let $ \Gamma $ be a context and $ M, A $ terms in an EPTS. We have
  \begin{itemize}
  \item If $ \Gamma~\WF $ then  $ \trans{\Gamma}_S~\textup{\texttt{well-formed}}_2 $
  \item If $  \Gamma \vdash M : A  $ then \begin{itemize}
    \item if $ A =s $ is a top-sort then $ \trans{\Gamma}_S \vdash_2 \trans{ M }_S : \ctb{U}~\senc{s}$
    \item else $ \trans{\Gamma}_S \vdash_2 \trans{ M }_S : \ctb{El}~\senc{s_A}~\trans{ A }_S$, where $ \Gamma \vdash A : s_A $
    \end{itemize}
  \end{itemize}      
\end{corollary}

In a sense, our notion of theory morphism allows us to embed a theory that is more fined grained into a theory that is less. For instance, to build our morphism $ \phi $, we map all the constants of the form $ \Prod{s_1}{s_2} $ to the same one. We then could try to build an inverse morphism $ \phi^{-1} $ to show conservativity of $ \trans{-}_S $, but this is not possible with our definition. Indeed, the same constant $ \ctb{Prod} $ should be sent into $ \Prod{s_1}{s_2} $ when it is applied to $ \senc{s_1}, \senc{s_2} $ and into into $ \Prod{s_3}{s_4} $ when it is applied to $ \senc{s_3}, \senc{s_4} $. However, in our definition the body of the implementation $ F_c $ only depends on the initial constant $ c $, and not on its arguments $ \vec{M} $.

Therefore, it is still an open problem for us to find a notion of morphism that would allows to build morphisms in both directions between $  (\Sigmaepts, \Repts ) $ and $ (\Sigmaepts^{S}, \Repts^S ) $, and then show the equivalence between the encodings. For the time being, in order to show conservativity  (and then adequacy)  of $ \trans{-}_S $ one basically  has to redo the work of Section \ref{sec:conservativity}.

We note nevertheless that our definition of morphism can have many other applications. For instance, if we consider two \Dedukti{} theories that express classical logic, one using the axiom of the excluded middle $ A \lor \neg A $ and the other using the double negation axiom $ A \Leftrightarrow \neg \neg A $, one could define morphisms in both directions in order to be able to transport proofs from a theory to another. It would suffice to map the constant representing the excluded middle $ \ctb{exm} $ to a proof of it $ F_{\ctb{exm}} $ which uses the double negation, and map the constant representing the double negation axiom $ \ctb{nnpp} $ to a proof of it $ F_{\ctb{nnpp}} $ which uses the excluded middle.

}

\section{The encoding in practice}
\label{sec:practice}

Our encoding satisfies nice theoretical properties, but when using it in practice it becomes quite annoying to have to explicit all the information needed in $ \app{s_1}{s_2} $ and $ \abs{s_1}{s_2} $. Worst, when performing translations from other systems where those parameters are not explicit we would then have to compute them during the translation. Thankfully, \Lambdapi\cite{lambdapi}, an implementation of \Dedukti, allows us to solve this by declaring some arguments as implicit, so they are only calculated internally.

Using the encoding of Figure \ref{theory-sort} we can  mark for instance the  arguments $ s_1, s_2, A $ of $ \ctb{Prod} $ as implicit. We can then also rename $ \ctb{Prod} $ into $ \ctb{\Pi'} $, $ \ctb{abs} $ into $ \ctb{\lambda'} $, $ \ctb{app} $ into $ \ctb{\sqbullet} $ and use  another \Lambdapi{} feature allowing to mark $ \ctb{\Pi'},\ctb{\lambda'} $ as quantifier and $ \ctb{\sqbullet} $ as infix left. This then allows us to represent $ \Pi x : A . B $ as $ \ctb{\Pi}' x : \ctb{El}~\trans{A}. \trans{B} $, $ \lambda x : A . B $ as $ \ctb{\lambda}' x : \ctb{El}~\trans{A}. \trans{B} $ and $ M~N $ as $ \trans{M} \ctb{\sqbullet} \trans{N} $. Using these notations, we can write terms in the encoding in a natural way, and we refer to \url{https://github.com/thiagofelicissimo/examples-encodigs} for a set of examples of this.

However, as \Dedukti~also aims to be used in practice for sharing real  libraries between proof assistants, we also tested how our approach copes with more practical scenarios. We provide in \url{https://github.com/thiagofelicissimo/encoding-benchmarking} a benchmark of Fermat's little theorem library in \Dedukti{}\cite{fermat}, where we compare the traditional encoding with an adequate version that applies the ideas of our approach\footnote{Because the underlying logic of the library is not a PTS, this encoding is not exactly the one we present here. However, it uses the same ideas discussed, and the same proof strategy to show adequacy applies.}. As we can see, the move from the traditional to the adequate version introduces a considerable performance hit. The standard \Dedukti{} implementation, which is our reference here, takes $ 16 $ times more time to typecheck the files. This is probably caused by the insertion of type parameters $ A $ and $ B $  in $ \abs{s_A}{s_B} $ and $ \app{s_A}{s_B} $, which are not needed in traditional encodings.

Nevertheless, \Dedukti{} is still able to typecheck our encoding within reasonable time, showing that our approach is indeed usable in practical scenarios, even if it is not the most performing one. Moreover, as our encoding is mainly intended to be used to check proofs, and not with interactive proof development, immediacy of the result is not essential and thus it can be reasonable to trade performance for better theoretical properties. Still, we plan in the future to look at techniques to improve our performances. In particular, using  more sharing in \Dedukti{} would probably reduce the time for typechecking, as the parameter annotations in $ \app{s_A}{s_B} $ and $ \abs{s_A}{s_B} $ carry a lot of repetition.

\section{Conclusion}
\label{sec:label}

By separating the framework's abstraction and application from the ones of the encoded system, we have proposed a new  paradigm for \Dedukti~encodings. Our approach offers much more well-behaved encodings, whose conservativity can be shown in a much more straightforward way and which feature adequacy theorems, something that was missing from traditional \Dedukti{} encodings. However, differently from the \ELF~approach, our encoding is also computational. Therefore, our method combines the adequacy of \ELF~encodings with the computational aspect of \Dedukti{} encodings. 

By decoupling the framework's $ \beta $ from the rewriting of the encoded system, our approach allows to show the expected properties of the encoding without requiring to show that the encoded system terminates. Indeed, our adequacy result concerns all functional EPTS, even non terminating ones, such as the one with  $ Type : Type $. This sets our work apart from \cite{dowek2007}, whose conservativity proof requires the encoded system to be normalizing.

This work opens many other directions we would like to explore. We believe that our technique can be extended to craft adequate and computational encodings of type theories with much more complex features, such as (co)inductive types, universe polymorphism, predicate subtyping and others. For instance, in the case of inductive types no type-level rewriting rules need to be added, thus  \thref{norm_beta_dk} would apply. Therefore, we could repeat the same technique of  normalizing only with $ \beta $ to show conservativity.

However, we would be particularly interested to see if we could take a general definition of type theories covering most of these features (maybe in the lines of \cite{bauer}). This would allow us to define a single encoding which could be applied to encode various features, and thus would saves us from redoing similar proofs multiple times.

\bibliography{ref}

\end{document}